\documentclass[aps,prd,superscriptaddress,onecolumn,showpacs,showkeys]{revtex4}
\usepackage{amssymb,amsmath,epsfig}
\usepackage{eurosym}
\usepackage{amsfonts}
\usepackage{array}
\usepackage{amsthm}
\usepackage{bm}
\usepackage{palatino}
\usepackage{changes}
\usepackage[colorlinks=true,
            linkcolor=blue,
            urlcolor=blue,
            citecolor=blue]{hyperref}
\usepackage{amssymb}
\usepackage{eurosym}
\usepackage{amsmath}
\usepackage{epsfig}
\usepackage{graphics}
\usepackage{color}
\usepackage{graphicx}
\usepackage{hyperref}

\newcommand{\be}{\begin{equation}}
\newcommand{\ee}{\end{equation}}
\newcommand{\ba}{\begin{eqnarray}}
\newcommand{\ea}{\end{eqnarray}}
\newcommand{\bal}{\begin{align}}
\newcommand{\eal}{\end{align}}

\newcommand{\bw}{\begin{widetext}}
\newcommand{\ew}{\end{widetext}}

\newcommand{\va}{\varepsilon}

\begin{document}

\title{Testing Generalized Einstein-Cartan-Kibble-Sciama Gravity using Weak Deflection Angle and Shadow Cast}

\author{Ali \"{O}vg\"{u}n}
\email{ali.ovgun@emu.edu.tr}
\homepage[]{https://www.aovgun.com}
\affiliation{Physics Department, Eastern Mediterranean
University, Famagusta, 99628 North Cyprus, via Mersin 10, Turkey}

\author{\.{I}zzet Sakall{\i}}
\email{izzet.sakalli@emu.edu.tr}
\affiliation{Physics Department, Eastern Mediterranean
University, Famagusta, 99628 North Cyprus, via Mersin 10, Turkey}

\begin{abstract}
In this paper, we use a new asymptotically flat and spherically symmetric solution in the generalized Einstein-Cartan-Kibble-Sciama (ECKS) theory of gravity to study the weak gravitational lensing and its shadow cast. To this end, we first compute the weak deflection angle of generalized ECKS black hole using the Gauss–Bonnet theorem in plasma medium and in vacuum. Next by using the Newman-Janis algorithm without complexification, we derive the rotating generalized ECKS black hole and in the sequel study its shadow. Then, we discuss the effects of the ECKS parameter on the weak deflection angle and shadow of the black hole. In short, the goal of this paper is to give contribution to the ECKS theory and look for evidences to understand how the ECKS parameter effects the gravitational lensing. Hence, we show that the weak deflection of black hole is increased with the increase of the ECKS parameter.
\end{abstract}
\date{\today}

\maketitle
\section{Introduction}

It is known that general relativity (GR) is the most successful and accurate gravitational theory at classical level \cite{Abbott:2016blz,Akiyama:2019cqa}. In GR, gravity is described as a geometric property of spacetime continuum; thus wise generalizing special theory relativity and Newton's law of universal gravitation. Furthermore, the background spacetime of GR is nothing but the Riemann manifold (represented as $\mathcal{V}_{4}$ ), which is torsionless. Let us recall that torsion is an antisymmetric part of the affine connection and it was first introduced by Cartan \cite{cartan}. 

There are various generalizations of Einstein's GR theory; one of which is the Einstein-Cartan theory that modifies the geometric structure of the manifold and relaxes the symmetric notion of affine connection. Einstein-Cartan theory is also known as $\mathcal{U}_{4}$ theory of gravitation \cite{cartan2,cartan3} in which the underlying manifold is not Riemannian. In fact, the non-Riemannian part of the spacetime is sourced by the spin density of matter such that the mass and spin both play the dynamical role. In particular, Cartan proved that the local Minkowskian structure of spacetime is not violated in the existence of torsion. So any manifold having torsion and curvature (with non-metricity $= 0 $ \cite{cartan4,cartan5}) can define physical spacetime very well. Since the works of Cartan \cite{cartan}, researchers have studied the theories of gravity on a Riemann-Cartan spacetime $\mathcal{U}_{4}$ over the last century \cite{Tecchiolli:2019hfe}. Among those studies, main framework of the Einstein-Cartan theory was laid down by Sciama and Kibble \cite{Sciama}; thus the theory is called the ECSK theory, which also takes into account effects from quantum mechanics. It not only provides a step towards quantum gravity but also leads to an alternative picture of the Universe. This variation of GR incorporates an important quantum property known as spin. In this theory, the curvature and the torsion are considered to be coupled with the energy and momentum and the intrinsic angular momentum of matter, respectively. The gravitational repulsion effect resulting from such a spinor-torsion coupling prevent the creation of spacetime singularities in the region with extremely high densities. Namely, spacetime torsion would only be significant, let alone noticeable, in the early Universe or in black holes \cite{Poplawski:2012ab}. In these extreme environments, spacetime torsion would manifest itself as a repulsive force that counters the attractive gravitational force coming from spacetime curvature. The repulsive torsion could create a "big bounce" like a compressed beach ball that snaps outward. The rapid recoil after such a big bounce could be what has led to our expanding Universe. The result of this recoil matches observations of the
Universe's shape, geometry, and distribution of mass \cite{sciama1,sciama2}. The torsion mechanism in effect suggests an incredible scenario: every black hole will create a new, baby Universe within. Therefore our own Universe may be inside a black hole that resides in another universe. Even as we can not see what is happening inside the black holes in the cosmos, any observers in the parent Universe will see what is happening within our cosmos.

The ECSK and GR theories offer indistinguishable predictions in the low density region, since the contribution from torsion to the Einstein equations is negligibly small. On the other hand, in the ECSK theory, the torsion field is not dynamic, because the torsion equation is an algebraic constraint rather than a partial differential equation, showing that the torsion field outside the distribution of matter vanishes since it can not disperse as a wave in spacetime. Recently, Chen et al \cite{Chen:2018szr} presented a new asymptotically flat and spherically symmetric solution in the generalized ECSK theory of gravity. They have also studied the wave dynamics of photon in the obtained geometry. It was found that the spacetime has three independent parameters which play role on the sharply photon sphere, deflection angle of light ray, and hence the gravitational lensing. In particular, there is a special case in the resulting spacetime that there is a photon sphere but no horizon. In that particular case, the angle of deflection of a light ray near the event horizon corresponds to a fixed value instead of diverging, which is not discussed in other spacetimes. Moreover, the strong gravitational lensing and how the spacetime parameters affect the coefficients in the strong field limit were also analyzed in \cite{Chen:2018szr}. It is worth noting that the gravitational lensing is a phenomenon of deflection of light rays in curved spacetimes. Gravitational lensing can provide us with many essential signatures on compact objects that can help us to detect black holes and test alternative gravitational theories. For bending angle having less than 1, weak deflection lensing has already been a gruelling in cosmology and gravitational physics \cite{izr1,izr2,izr3}. With a key ingredient that a photon might be able to go around a black hole, by at least one loop, strong deflection lensing can form the shadow and relativistic images. The Event Horizon Telescope imaged the shadow of M87* with measured diameter of 42 microarcsecond \cite{izr4,izr5,izr6,izr7,izr8,izr9}. Recently, attempts have been made to directly visualize the shadow of Sagittarius A* \cite{izr10}, the supermassive black hole in the Galactic Center, and likely observational results will soon be revealed. Thanks to the relativistic images that we will be able to better understand the nature of black holes and distinguish their different types in the near future \cite{Chen:2018szr,Blake:2020mzy,Joachimi:2020abi,Monteiro-Oliveira:2020yfy,Yu:2020agu,Foxley-Marrable:2020ckf,Tu:2019vcj,Qi:2019zdk}. Since Eddington's first gravitational lensing observation \cite{edd}, numerous works have been published on the gravitational lensing for black holes, wormholes, celestial strings, and other compact objects. For example: Bartelmann and Schneider review theory and applications of weak gravitational lensing in \cite{Bartelmann:1999yn}, Bozza studies an analytic method to discriminate among different types of black holes on the ground
of their strong field gravitational lensing properties \cite{Bozza:2002zj}, Tsukamoto et. al show that it is possible to distinguish between slowly rotating Kerr-Newmann black holes and the Ellis wormholes with their Einstein-ring systems \cite{Tsukamoto:2012xs}.  Moreover, Aazami et. al develop an analytical theory of quasi-equatorial lensing by Kerr black holes \cite{Aazami:2011tw}. Virbhadra et. al, show that the lensing features are qualitatively similar for the Schwarzschild black holes, weakly naked, and marginally strongly naked
singularities \cite{Virbhadra:2007kw}. Ishak and Rindler study some recent developments concerning the effect
of the cosmological constant on the bending of light \cite{Ishak:2010zh}. Keeton and Petters provide the new formalism for computing corrections to lensing observables for static,
spherically symmetric gravity theories \cite{Keeton:2005jd}. Wei et. al study the strong
gravitational lensing by the asymptotic flat charged Eddington-inspired Born-Infeld black hole \cite{Wei:2014dka}. Moreover, Iyer and Petters  obtain an invariant series for
the strong-deflection bending angle that extends beyond the standard logarithmic deflection term
used in the literature \cite{Iyer:2006cn}. Also there are various works in literature related to gravitational lensing \cite{Gibbons:2011rh}-\cite{Sharif:2015kna}.

Gibbons and Werner discovered a very successful approach to obtain the angle of light deflection from non-rotating asymptotically flat spacetimes \cite{Gibbons:2008rj}. In the sequel, Werner \cite{Werner2012} generalized the method to stationary space times.  Gibbons and Werner's mechanism has been applied to many curved spacetimes over the last 10 years and very successful results have been accomplished \cite{Crisnejo:2018uyn,Jusufi:string17,Jusufi&Ali:Teo,Jusufi&Ali:string,Jusufi:RB,Jusufi:monopole,Ali:wormhole,Ali:strings,Ali:BML,Javed1,Javed2,Javed3,Javed4,Sakalli2017,Goulart2018,Leon2019,LZ20201,zhu2019,444,445,446,447,448,449,450,1765621,CGJ2019,Jusufi:mp,LHZ2020,ISOA2016,IOA2017,OIA2017,OIA2018,OIA2019,OA2019,LA2020,LJ2020,LZ20202,Arakida2018,Takizawa2020,Gibbons2016,Islam:2020xmy,Pantig:2020odu,Takizawa:2020egm,Kumar:2020hgm,Tsukamoto:2020uay,Crisnejo:2019xtp}. Their method is mainly based on the Gauss-Bonnet theorem and the optical geometry of the black hole's spacetime, where the source and receiver are located at IR regions. This approach was also extended to the finite distances \cite{ISOA2016,IOA2017,OIA2017,OIA2018,OIA2019,OA2019,Arakida2018}.

Another important event for the black hole is their shadow cast: a two dimensional dark zone in the celestial sphere caused by the strong gravity of the black hole firstly studied by Synge in 1966 and then Luminet find the angular radius for the shadow \cite{synge,luminet}. Material, such as gas, dust and other stellar debris that has come close to a black hole but outside of the event horizon to fall into it, forms a flattened band of spinning matter around the event horizon called the accretion disk. Event horizon of black hole is invisible,  however, this accretion disk can be seen, because the spinning particles are accelerated to tremendous speeds by the huge gravity of the black hole, by releasing heat and powerful x-rays and gamma rays out into the universe as they smash into each other \cite{cunha,Cunha:2018acu,Falcke:1999pj,Tremblay:2016ijg}. Moreover, this accreting matter heats up
through viscous dissipation and radiate light in various frequencies such as radio waves which can be detected through the
radio telescopes \cite{Shen:2005cw,Huang:2007us,Johannsen:2015mdd}. Namely, the dark region in the center is termed the black hole's “shadow”; this is the collection of paths of photons that did not escape, but were instead captured by the black hole.
\cite{Cunha:2018gql}. We can say that this shadow is actually an image of
the event horizon. The center of galaxies is a playground of a gigantic black holes. Because of
the gravitational lens effect, the background would have cast a shade larger
than its horizon size. The size and shape of this shadow can be
calculated and visualized, respectively. The radius of the black hole's shadow calculated as $r_{shadow}=\sqrt{27}M=5.2M$ 
 \cite{bardeen,Chandra}. After that, the shadows of black holes (and also the wormholes) have been investigated by several authors. For example: Hioki and Maeda provided a method to determine the spin parameter and the inclination angle by observing the apparent
shape of the shadow \cite{Hioki:2009na}, Johannsen and Psaltis verified the no-hair theorem by using the black hole shadow \cite{Johannsen:2010ru}. Nedkova et. al studied the shadow of a rotating traversable wormhole \cite{Nedkova:2013msa}. Amarilla and Eiroa analyzed the shadow of a Kaluza-Klein rotating dilaton black hole \cite{Amarilla:2013sj}. In the same line of thought, Abdujabbarov et al. studied the shadow of Kerr-Taub-NUT black hole \cite{Abdujabbarov:2012bn}. Next, Grenzebach et al. studied photon regions and shadows of the Kerr-Newman-NUT black holes with a cosmological constant \cite{Grenzebach:2014fha}. Then, Johannsen et. al tested the general relativity with the shadow size of Sgr A* \cite{Johannsen:2015hib} and Giddings investigated the possible quantum effects arising from the black hole shadows \cite{Giddings:2014ova}. Following this, the shadows of rotating non-Kerr and Einstein-Maxwell-dilaton-axion black holes were obtained by Atamurotov et al. \cite{Atamurotov:2013sca} and Wei and Liu \cite{Wei:2013kza}, respectively. Then, the shadow of Gravastar was investigated by Sakai et. al \cite{Sakai:2014pga}. In the sequel, Perlick et. al studied the influence of a plasma on the shadow of a spherically symmetric black hole \cite{Perlick:2015vta}. Hereupon, Abdujabbarov et. al studied the coordinate-independent characterization of a black hole shadow \cite{Abdujabbarov:2015xqa} and Tinchev and Yazadjiev revealed the possible imprints of cosmic strings in the shadows of galactic black holes \cite{Tinchev:2013nba}. Moreover, Wang et. al showed the chaotic behaviour of the shadow for a non-Kerr rotating compact object having quadrupole mass moment \cite{Wang:2018eui}. While Amarilla et. al  \cite{Amarilla:2010zq} studied the shadow of a rotating black hole in the extended Chern-Simons modified gravity, Yumoto et. al obtained the shadows of multi black holes \cite{Yumoto:2012kz}. Remarkably, Takahashi studied the shadows of charged spinning black holes \cite{Takahashi:2005hy} and Papnoi et. al obtained the shadow of five-dimensional rotating Myers-Perry black holes \cite{Papnoi:2014aaa}. Recently, ~\"{O}vg\"{u}n et al. have studied the shadow cast of non-commutative black holes in Rastall gravity \cite{Ovgun:2019jdo}. Furthermore, there are also various studies on the black hole shadow such as for the tilted black holes \cite{Dexter:2012fh}, for the modified gravity black holes \cite{Moffat:2015kva}, for the parametrized axisymmetric black holes \cite{Younsi:2016azx}, for the Kerr-like black holes \cite{Johannsen:2015qca}, which constraints on a charge in the Reissner–Nordstr\"{o}m metric for the
black hole at the galactic center \cite{Zakharov:2014lqa}, for boson stars \cite{Cunha:2016bjh}, for holographic reconstruction \cite{Freivogel:2014lja}, for Kerr black holes with and without scalar hair \cite{Cunha:2016bpi}, for wormholes \cite{Ohgami:2015nra}, for evaluate black hole parameters and a dimension of spacetime \cite{Zakharov:2011zz}, for black holes in Einsteinian cubic gravity \cite{Hennigar:2018hza}, and more which can be seen in Refs. \cite{Pu:2016qak}-\cite{Bambi:2008jg}. %

The aim of this work is to study the deflection angle provided by the regular black holes obtained in the ECSK theory using the Gauss-Bonnet theorem and thus to investigate the effect of torsion on the gravitational lensing. Since the torsion could be the source of "dark energy" \cite{Poplawski:2012bw}, a
mysterious form of energy that permeates all of space and increases the rate of expansion of the Universe, we do want to contribute to the work on the gravitational lensing effects of dark matter. 
The paper is organized as follows. In Sec. II,  we briefly review the black hole obtained in the ECSK gravitational theory \cite{Chen:2018szr}. We compute the deflection angle by the ECSK black hole using the Gibbons and Werner's approach (i.e., via the Gauss-Bonnet theorem)
in the weak field regime in Sec. III. Section IV is devoted to the study of the deflection of light by the ECSK black hole in a plasma
medium. We then introduce the rotating generalized ECKS black hole and study its shadow cast in Secs. V and VI, respectively. Finally,
we present our conclusions in Sec. VII.

\section{Black holes in the generalized
Einstein-Cartan-Kibble-Sciama gravity}
In this section, we briefly review the black hole solution in generalized ECKS theory. The action for the generalized ECKS theory with Ricci scalar $\tilde{R}$ and torsion scalar $\mathcal{T}$ is given by \cite{Chen:2018szr}:
\begin{eqnarray}
S&=&\int d^4x \sqrt{-g}\bigg[-\frac{1}{16\pi G}\bigg(\tilde{R}+\tilde{R}\mathcal{T}\bigg)\bigg],
\label{action}
\end{eqnarray}
where
\begin{eqnarray}
\tilde{R}&=& R+\frac{1}{4}Q_{\alpha\beta\gamma}Q^{\alpha\beta\gamma}
+\frac{1}{2}Q_{\alpha\beta\gamma}Q^{\beta\alpha\gamma}+
Q^{\alpha\;\beta}_{\;\alpha}Q^{\gamma}_{\;\beta\gamma}+2Q^{\alpha\;\beta}_{\;\alpha;\beta},
\nonumber\\
\mathcal{T}&=&a_1Q_{\alpha\beta\gamma}Q^{\alpha\beta\gamma}
+a_2Q_{\alpha\beta\gamma}Q^{\alpha\gamma\beta}+
a_3Q^{\alpha\;\beta}_{\;\alpha}Q^{\gamma}_{\;\gamma\beta}.
\end{eqnarray}
Note that $\tilde{R}$ and $R$ stand for Riemann curvature scalar related with general affine connection $\tilde{\Gamma}^{\alpha}_{\;\mu\nu}$ and Levi-Civita Christoffel connection $\Gamma^{\alpha}_{\;\mu\nu}$, respectively.
Moreover, tensor $Q^{\alpha}_{\;\mu\nu}$ shows the torsion of spacetime defined by
$Q^{\alpha}_{\;\mu\nu}=\tilde{\Gamma}^{\alpha}_{\;\mu\nu}-\tilde{\Gamma}^{\alpha}_{\;\nu\mu}.$ The affine connection $\tilde{\Gamma}^{\alpha}_{\;\mu\nu}$ can be calculated by using the Christoffel connection $\Gamma^{\alpha}_{\;\mu\nu}$:
$\tilde{\Gamma}^{\alpha}_{\;\mu\nu}=\Gamma^{\alpha}_{\;\mu\nu}+K^{\alpha}_{\;\mu\nu},$
with the contorsion tensor:
$K^{\alpha}_{\;\mu\nu}=\frac{1}{2}\bigg[Q^{\alpha}_{\;\mu\nu}-Q^{\;\alpha}_{\mu\;\nu}
-Q^{\;\alpha}_{\nu\;\mu}\bigg].$

The static spherically symmetric spacetime in the generalized ECKS theory of gravity (\ref{action}) is given by \cite{Chen:2018szr}
\begin{eqnarray}
ds^2 = -f(r)dt^2 + \frac{dr^2}{g(r)} + r^2 (d\theta^2
+\sin^2\theta d\phi^2), \label{metr}
\end{eqnarray}
where $H$ and $F$ are only functions of the polar coordinate $r$:

\begin{eqnarray}
\label{sol1}
f(r)&=&1-\frac{2 \gamma m}{r}+\frac{q^2}{r^2},
\nonumber\\
g(r)&=&\frac{1}{(\gamma^2 m^2-q^2)^2}\bigg[\gamma (\gamma-1)m^2+\frac{(1-\gamma)q^2m+(\gamma m^2-q^2)\sqrt{r^2-2 \gamma m r+q^2}}{r}\bigg]^2,
\end{eqnarray}
where $m$, $q$, and $\gamma$ are constants.
It is noted that this solution (\ref{sol1}) is asymptotically flat because $f(r)$ and $g(r)$ go to $1$ when $r$ approaches to spatial infinity. Moreover, the solution (\ref{sol1}) reduces to the Reissner-Nordstr\"{o}m black hole for $\gamma=1$ and thus to the Schwarzschild black hole ($q=0$). Remarkably, this solution corresponds to a scalar-tensor wormhole if the parameter replacement $q^2\rightarrow-\beta$, $\gamma m\rightarrow m$, and $(\gamma m^2-q^2)/(\gamma(\gamma-1)m^2)\rightarrow \eta$.
%H=f  and F=g
The event horizon is found to be [i.e., upon the condition of $g(r)=0$]
\begin{eqnarray}\label{EH1}
r_H=\gamma m+\sqrt{\gamma^2m^2-q^2}.
\end{eqnarray}
Thus, the surface gravity \cite{myWald} can be computed as
\begin{eqnarray}
\kappa=\frac{1}{2}\sqrt{\frac{g^{rr}}{-g_{tt}}}\frac{-dg_{tt}}{dr}\bigg|_{r=r_H}=
\frac{1}{2}\sqrt{\frac{g(r)}{f(r)}}\frac{d f(r)}{dr}\bigg|_{r=r_H}=\frac{\gamma m^2-q^2}{\sqrt{\gamma^2m^2-q^2}(\gamma m+\sqrt{\gamma^2m^2-q^2})^2}.
\end{eqnarray}

ECKS parameter $\gamma$ can have a major impact on the path of photons moving in the cosmos. We shall try to take this effect into account in the upcoming sections.

\section{Deflection angle of photons by black hole in the generalized
Einstein-Cartan-Kibble-Sciama gravity using Gauss-Bonnet Theorem}

In this section, we study the weak gravitational lensing in the background of the ECKS black hole by using the Gauss–Bonnet theorem. To do so, we first obtain the optical metric within the equatorial plane $\theta =\pi /2$:
\begin{equation}
\mathrm{d}t^{2}=\frac{1}{f(r)g(r)}\mathrm{d}r^2+\frac{r^2}{f(r)}\mathrm{d}\varphi^2.
\end{equation}

Afterwards, we calculate the corresponding Gaussian optical curvature $K=\frac{R_{icciScalar}}{2}$: 
\begin{eqnarray}
  \mathcal{K} & = &\,{\frac {2\,f \left( r \right) g \left( r \right)  \left( {\frac {
{\rm d}^{2}}{{\rm d}{r}^{2}}}f \left( r \right)  \right) r-2\,g
 \left( r \right)  \left( {\frac {\rm d}{{\rm d}r}}f \left( r \right) 
 \right) ^{2}r+f \left( r \right)  \left( r{\frac {\rm d}{{\rm d}r}}g
 \left( r \right) +2\,g \left( r \right)  \right) {\frac {\rm d}{
{\rm d}r}}f \left( r \right) -2\, \left( f \left( r \right)  \right) ^
{2}{\frac {\rm d}{{\rm d}r}}g \left( r \right) }{2f \left( r \right) r}
},
\end{eqnarray}

which reduces to the following form in the weak field limit approximation:

\begin{align}\label{Curvature1}
\begin{split}
\mathcal{K} \approx 2\,{\frac {{m}^{2}{q}^{2}}{{r}^{6}}}-9/2\,{\frac {m{q}^{2}}{{r}^{5}}}-
{\frac {m}{{r}^{3}}}+3\,{\frac {{q}^{2}}{{r}^{4}}}-1/2\,{\frac {{m}^{2
}{q}^{2}\gamma}{{r}^{6}}}+3\,{\frac {{m}^{2}\gamma}{{r}^{4}}}-3/2\,{\frac {m\gamma{q}
^{2}}{{r}^{5}}}-{\frac {m\gamma}{{r}^{3}}}-3/2\,{\frac {{m}^{2}{q}^{2}{\gamma}^{
2}}{{r}^{6}}}.
\end{split}
\end{align}
To calculate the weak deflection angle, we define a non-singular region $\mathcal{D}_{R}$ with boundary $\partial
\mathcal{D}_{R}=\gamma _{\tilde{g}}\cup C_{R}$ and then apply the Gauss–Bonnet theorem \cite{Gibbons:2008rj}:
\begin{equation}
\iint\limits_{\mathcal{D}_{R}}\mathcal{K}\,\mathrm{d}S+\oint\limits_{\partial \mathcal{%
D}_{R}}\kappa \,\mathrm{d}t+\sum_{i}\theta _{i}=2\pi \chi (\mathcal{D}_{R}),
\end{equation}
where $\kappa $ is for the geodesic curvature. Note that $\theta_{i}$ is the exterior angle at the $i^{th}$ vertex. One can choose the region which is outside of the light ray and the Euler characteristic number $%
\chi (\mathcal{D}_{R})=1$.  Then, one can calculate the geodesic curvature $\kappa =\tilde{g}\,\left(\nabla _{\dot{%
\gamma}}\dot{\gamma},\ddot{\gamma}\right)$ using the unit speed condition $\tilde{g}(\dot{\gamma},\dot{%
\gamma})=1$, with $\ddot{\gamma}$ the unit acceleration vector. When $R\rightarrow \infty $, two jump angles ($\theta _{\mathcal{O}}$, $%
\theta _{\mathcal{S}}$) is $\pi /2$. Then, the Gauss–Bonnet theorem can be written as follows:
\begin{equation}
\iint\limits_{\mathcal{D}_{R}}\mathcal{K}\,\mathrm{d}S+\oint\limits_{C_{R}}\kappa \,%
\mathrm{d}t\overset{{R\rightarrow \infty }}{=}\iint\limits_{\mathcal{D}%
_{\infty }}\mathcal{K}\,\mathrm{d}S+\int\limits_{0}^{\pi +\alpha}\mathrm{d}\varphi
=\pi.
\end{equation}

Note that $\kappa (\gamma _{\tilde{g}})=0$. Since $\gamma _{\tilde{g}}$ is a
geodesic, we have
\begin{equation}
\kappa (C_{R})=|\nabla _{\dot{C}_{R}}\dot{C}_{R}|,
\end{equation}
in which  $C_{R}:=r(\varphi)=R=\text{const}$. The radial part is calculated as follows:
\begin{equation}
\left( \nabla _{\dot{C}_{R}}\dot{C}_{R}\right) ^{r}=\dot{C}_{R}^{\varphi
}\,\left( \partial _{\varphi }\dot{C}_{R}^{r}\right) +\tilde{\Gamma} _{\varphi
\varphi }^{r}\left( \dot{C}_{R}^{\varphi }\right) ^{2}. \label{12}
\end{equation}
The first term in the above equation vanishes and second term is found by using the unit speed condition. Then, $\kappa $ is obtained as follows:

\begin{eqnarray}\notag
\lim_{R\rightarrow \infty }\kappa (C_{R}) &=&\lim_{R\rightarrow \infty
}\left\vert \nabla _{\dot{C}_{R}}\dot{C}_{R}\right\vert. \notag \\
&\rightarrow &\frac{1}{R}. 
\end{eqnarray}%

At the large limits of the radial distance, one gets
\begin{eqnarray}
\lim_{R\rightarrow \infty } \mathrm{d}t&\to & \left(R\right) \, \mathrm{d}\varphi.  
\end{eqnarray}%

Combining the last two equations, one can get $
\kappa (C_{R})\mathrm{d}t= \mathrm{d}\,\varphi
$. Using the straight light approximation, we find $r=b/\sin \varphi$, where $b$ is the impact parameter. Hence it is shown that Gauss–Bonnet theorem reduces to this form for calculating deflection angle \cite{Gibbons:2008rj}:
\begin{eqnarray}\label{int0}
\alpha=-\int\limits_{0}^{\pi}\int\limits_{\frac{b}{\sin \varphi}}^{\infty}\mathcal{K}\mathrm{d}S.
\end{eqnarray}

Solving the above integral with the Gaussian curvature, the weak deflection angle up to the second order terms is found as follows:

\begin{equation}\label{deflection}
\begin{split}
\alpha \approx 2\,{\frac {m\gamma}{b}}-\,{\frac {3{q}^{2}\pi}{4{b}^{2}}}+2\,{\frac {m}{b}
}.\end{split}
\end{equation} 

\begin{figure}[htbp]
\centering
\begin{minipage}[t]{0.44 \textwidth}
    \includegraphics[scale=0.70]{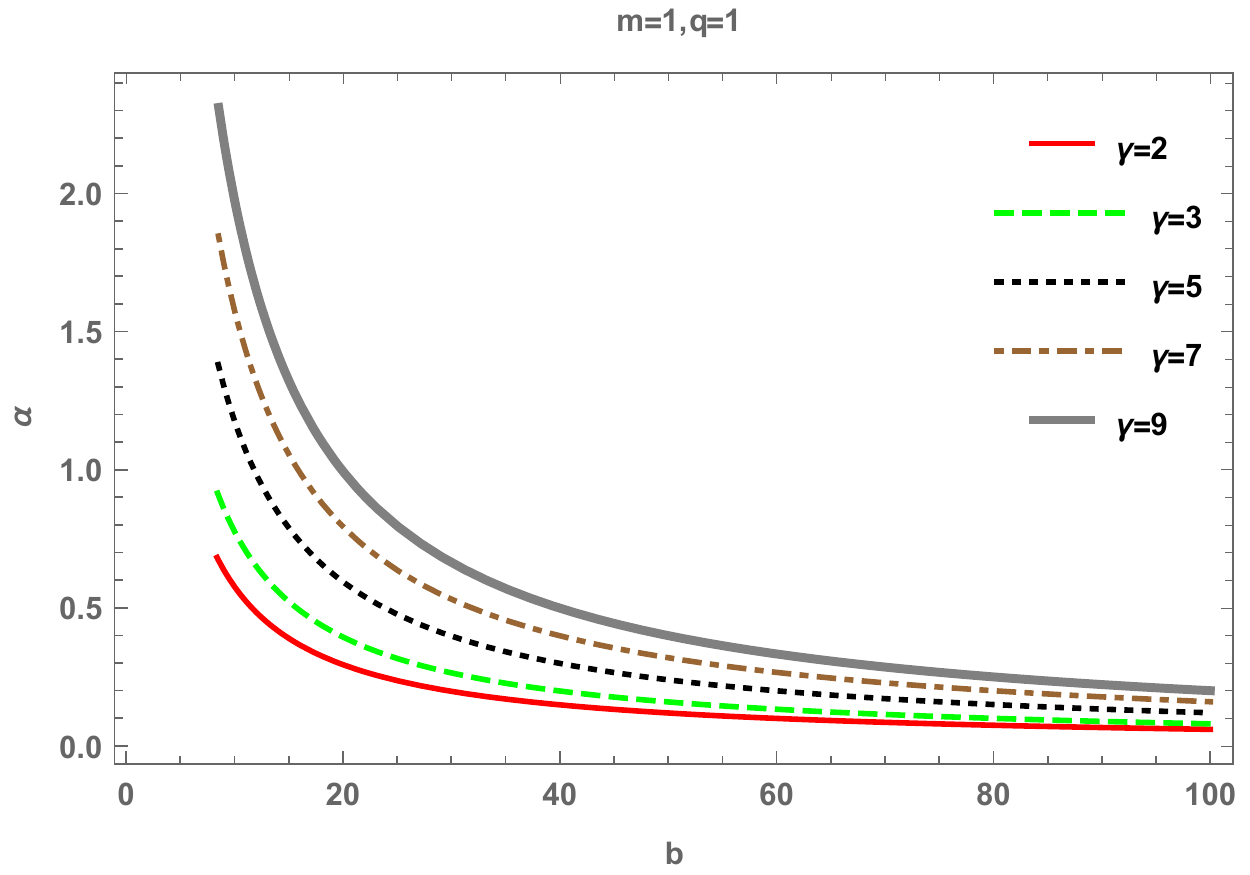}
        \caption{ $\alpha$ versus $b$ to see the influence of $\gamma$ parameter on weak deflection angle.}
         \label{fig:00}
     \end{minipage}
\hfill
\begin{minipage}[t]{0.47\textwidth}
        \includegraphics[scale=0.70]{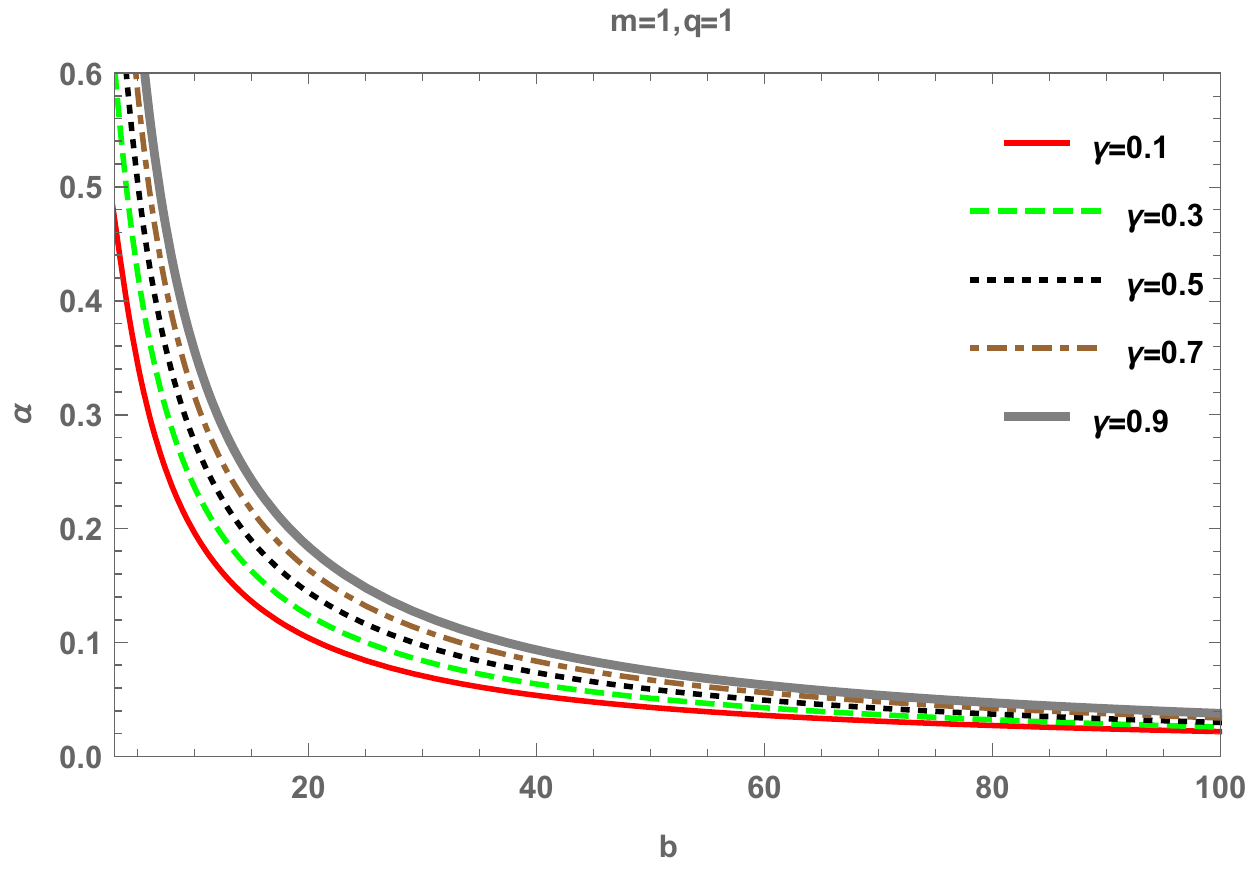}
    \caption{ $\alpha$ versus $b$ to see the influence of $\gamma$ parameter on weak deflection angle.}
       \label{fig:01}
       \end{minipage}
\end{figure}

It is obvious that the ECKS parameter $\gamma$ increases with the weak deflection angle as seen in Figs. (1-2).

\section{Deflection angle of photons in plasma medium by black hole in the generalized
Einstein-Cartan-Kibble-Sciama gravity}

In this section, we study the effect of a plasma medium on the weak deflection angle by generalized ECKS black holes. The refractive index of the cold plasma medium $n(r)$ is obtained as \cite{Crisnejo:2018uyn}:
\begin{equation}
n(r)=\sqrt{1-\frac{\omega_{e}^{2}}{\omega^{2}_{\infty}} \bigg(\frac{f(r)}{f_{\infty}}\bigg)},
\end{equation}
in which  $\omega_{e}$ is the electron plasma frequency and $\omega_{\infty}$ is the photon frequency measured by an observer at infinity. Afterwards, we calculate the corresponding optical metric:

\begin{equation}
d \sigma ^ { 2 } = g _ { i j } ^ { \mathrm { opt } } d x ^ { i } d x ^ { j } =\frac{n^{2}(r)}{f(r)}\left(\frac{d r^{2}}{g(r)}+r^2 d \varphi^{2}\right).
\end{equation}

The Gaussian curvature for the above optical metric is calculated as follows: 
\begin{equation}
\mathcal{K}\approx
-{\frac {\gamma m{\omega_{e}}^{2}}{{\omega_{\infty}}^{2}{r}^{3}}}-{
\frac {\gamma m}{{r}^{3}}}-2\,{\frac {m{\omega_{e}}^{2}}{{\omega_{\infty}}^{2}{r}^{3}}}-{\frac {m}{{r}^{3}}}+5\,{\frac {{q}^{2}{\omega_{e}}^{2}}{{\omega_{\infty}}^{2}{r}^{4}}}+3\,{\frac {{q}^{2}}{{r}
^{4}}},
\end{equation}

We have also
\begin{equation}
\frac{d\sigma}{d\varphi}\bigg|_{C_{R}}=
n ( R ) \left( \frac { r^2 } { f ( R ) } \right) ^ { 1 / 2 },
\end{equation}
which has the following limit:
\begin{equation}
\lim_{R\to\infty} \kappa_g\frac{d\sigma}{d\varphi}\bigg|_{C_R}=1\,.
\end{equation}
At spatial infinity, $R\to\infty$, and by using the straight light approximation $r=b/\sin\varphi$, the Gauss-Bonnet theorem reduces to \cite{Gibbons:2008rj,Crisnejo:2018uyn}:
\begin{equation}
\lim_{R\to\infty} \int^{\pi+\alpha}_0 \left[\kappa_g\frac{d\sigma}{d\varphi}\right]\bigg|_{C_R}d\varphi =\pi-\lim_{R\to\infty}\int^\pi_0\int^R_{\frac{b}{\sin\varphi}}\mathcal{K} dS.
\end{equation}
We calculate the weak deflection angle in the weak limit approximation as follows:
\begin{equation}\label{eq:perf}
\alpha \approx
-5/4\,{\frac {{q}^{2}{\omega_{e}}^{2}\pi}{{\omega_{\infty}}^{2}
{b}^{2}}}-3/4\,{\frac {{q}^{2}\pi}{{b}^{2}}}+2\,{\frac {\gamma m{\omega_{e}}^{2}}{b{\omega_{\infty}}^{2}}}+2\,{\frac {\gamma m}{b}}+4\,{\frac 
{m{\omega_{e}}^{2}}{b{\omega_{\infty}}^{2}}}+2\,{\frac {m}{b}}.
\end{equation}
Hence, we show that the photon rays move in a medium of homogeneous plasma. Note that $\omega_e/\omega_\infty\to 0$, Eq. \eqref{eq:perf} reduces to Eq. \eqref{deflection}, and thus the effect of the plasma is terminated. Moreover, the solution (\ref{eq:perf}) with $\omega_e/\omega_\infty\to 0$ reduces to deflection angle of Reissner-Nordstr\"{o}m black hole for $\gamma=1$ and it also corresponds to a scalar-tensor wormhole if the parameter replacement $q^2\rightarrow-\beta$, $\gamma m\rightarrow m$, and $(\gamma m^2-q^2)/(\gamma(\gamma-1)m^2)\rightarrow \eta$.

\section{Rotating Generalized ECKS Black hole}
Here, we briefly review the method of Newman-Janis without complexification presented by Azreg-Ainou \cite{Azreg-Ainou:2014pra} for transforming static spacetimes to stationary spacetimes. The generic four dimensional static and spherically symmetric spacetime can be written as follows:
\begin{equation}
ds^2=-f(r) dt^2+\frac{dr^2}{g(r)}+h(r)\left(d\theta^2+\sin^2\theta d\phi^2\right).
\end{equation}
First, the above metric is transformed into the advanced null Eddington-Finkelstein (EF) coordinates $(u,r,\theta,\phi)$ by defining the transformation of $du=dt-\frac{dr}{\sqrt{fg}}.$ Afterwards, the metric in EF coordinates takes the following form \cite{Shaikh:2019fpu}:
\begin{equation}
ds^2=-f du^2-2\sqrt{\frac{f}{g}}dudr+h\left(d\theta^2+\sin^2\theta d\phi^2\right).
\end{equation}
Secondly, one should write the inverse metric $g^{\mu\nu}$ with a null tetrad $Z_\alpha^\mu=(l^\mu,n^\mu,m^\mu,\bar{m}^\mu)$ using the form of
$g^{\mu\nu}=-l^\mu n^\nu -l^\nu n^\mu +m^\mu \bar{m}^\nu +m^\nu \bar{m}^\mu,
$
in which $\bar{m}^\mu$ is the complex conjugate of $m^\mu$. Moreover, the tetrad vectors must satisfy the following relations:
$
l_\mu l^\mu = n_\mu n^\mu = m_\mu m^\mu = l_\mu m^\mu = n_\mu m^\mu =0,
$ and 
$
l_\mu n^\mu = - m_\mu \bar{m}^\mu =-1.
$
Using the above conditions, null tetrads become
\begin{equation}
l^\mu=\delta^\mu_r, \, n^\mu=\sqrt{\frac{g}{f}}\delta^\mu_u-\frac{g}{2}\delta^\mu_r, \, m^\mu=\frac{1}{\sqrt{2h}}\left(\delta^\mu_\theta+\frac{i}{\sin\theta}\delta^\mu_\phi\right).
\end{equation}
Afterwards, using the transformation $
r\rightarrow r'=r+ia\cos\theta, \hspace{0.3cm} u\rightarrow u'=u-ia\cos\theta,$
with the spin parameter $a$. 

Third step is to use complexification which is proposed by the Azreg-Ainou\cite{Azreg-Ainou:2014pra}. Using this method: the metric functions $f(r)$, $g(r)$ and $h(r)$  transform to $F = F(r, a, \theta)$, $ G = G(r, a, \theta)$ and $H = H(r, a, \theta)$, respectively. Then we can write the null tetrads in terms of new metric functions:
\begin{equation}
l'^\mu=\delta^\mu_r, \quad n'^\mu=\sqrt{\frac{G}{F}}\delta^\mu_u-\frac{G}{2}\delta^\mu_r,
\end{equation}
\begin{equation}
m'^\mu=\frac{1}{\sqrt{2H}}\left(ia\sin\theta(\delta^\mu_u-\delta^\mu_r)+\delta^\mu_\theta+\frac{i}{\sin\theta}\delta^\mu_\phi\right).
\end{equation}
Then the inverse metric become:
\begin{equation}
g^{\mu\nu}=-l'^\mu n'^\nu -l'^\nu n'^\mu +m'^\mu \bar{m}'^\nu +m'^\nu \bar{m}'^\mu.
\end{equation}
Hence, we can write the new spacetime metric in the EF coordinates as follows
\begin{eqnarray}\notag
ds^2&=&-Fdu^2-2\sqrt{\frac{F}{G}}dudr+2a\sin^2\theta\left(F-\sqrt{\frac{F}{G}}\right)du d\phi+ 2a\sqrt{\frac{F}{G}}\sin^2\theta drd\phi+H d\theta^2 \\
&+&\sin^2\theta\left[H+a^2\sin^2\theta\left(2\sqrt{\frac{F}{G}}-F\right)\right]d\phi^2.
\label{eq:null_coordinate_metric_1}
\end{eqnarray}
After that, we transform the above metric to the Boyer-Lindquist (BL) coordinates using  $du=dt'+\va(r)dr$ and $ d\phi=d\phi'+\chi(r) dr$:
\begin{equation}
\va(r)=-\frac{k(r)+a^2}{g(r)h(r)+a^2},
\end{equation}
\begin{equation}
\chi(r)=-\frac{a}{g(r)h(r)+a^2},
\end{equation}
\begin{equation}
k(r)=\sqrt{\frac{g(r)}{f(r)}}h(r).
\end{equation}

Fixing some terms, one can get the unknown functions $F$, $G$, and $H$: 
\begin{equation}
F(r)=\frac{(g(r) h(r)+a^2 \cos^2\theta) H}{(k(r)+a^2 \cos^2\theta)^2},
\end{equation}
\begin{equation}
G(r)=\frac{g(r) h(r)+a^2 \cos^2\theta}{H}.
\end{equation}
Hence, the stationary spacetime metric is found as follows:
\begin{equation}
d s^{2}=\frac{H}{k+a^{2} \cos ^{2} \theta}\left[\left(1-\frac{\sigma}{k+a^{2} \cos ^{2} \theta}\right) d t^{2}\right. \\
-\frac{k+a^{2} \cos ^{2} \theta}{\Delta} d r^{2}+\frac{2 a \sigma \sin ^{2} \theta}{k+a^{2} \cos ^{2} \theta} d t d \phi
\\\notag
\end{equation}
\begin{equation}
-\left(k+a^{2} \cos ^{2} \theta\right) d \theta^{2} \left.-\frac{\left[\left(k+a^{2}\right)^{2}-a^{2} \Delta \sin ^{2} \theta\right] \sin ^{2} \theta}{k+a^{2} \cos ^{2} \theta} d \phi^{2}\right]
.
\end{equation}
in which $\sigma(r) \equiv k-g h$, $  \quad \Delta(r) \equiv g h+a^{2}$, and $ k \equiv \sqrt{\frac{g(r)}{f(r)}} h(r).$
Here $h(r)=r^2$ and  the other metric functions of the rotating black hole solution are given by
\begin{equation}
    \sigma(r)=2\,\gamma mr+\sqrt {-{\frac { \left( {\gamma}^{2}{m}^{2}-{q}^{2} \right) ^{2}
 \left( 2\,\gamma mr-{q}^{2}-{r}^{2} \right) }{ \left(  \left( \gamma{m}^{2}-{q}^
{2} \right) \sqrt {-2\,\gamma mr+{q}^{2}+{r}^{2}}+m \left( \gamma-1 \right) 
 \left( \gamma mr-{q}^{2} \right)  \right) ^{2}}}}{r}^{2}-{q}^{2}-{r}^{2}
\end{equation}
\begin{equation}
    \quad \Delta(r)= -2\,\gamma mr+{a}^{2}+{q}^{2}+{r}^{2},
\end{equation}
and
\begin{equation}
   k=\sqrt {-{\frac { \left( {\gamma}^{2}{m}^{2}-{q}^{2} \right) ^{2} \left( 2\,
\gamma mr-{q}^{2}-{r}^{2} \right) }{ \left(  \left( \gamma {m}^{2}-{q}^{2}
 \right) \sqrt {-2\,\gamma mr+{q}^{2}+{r}^{2}}+m \left( \gamma-1 \right)  \left( 
\gamma mr-{q}^{2} \right)  \right) ^{2}}}}{r}^{2}.
\end{equation}

The horizons of the rotating generalized ECKS black hole are obtained from the condition of $g^{rr}=0$ (one can see that $g_{rr}=\frac{H}{\quad \Delta(r)}$) . In the non-rotating limit, $h=\lim _{a \rightarrow 0} H$, the horizons are easily obtained from $\Delta(r) = 0$.

\section{Shadow cast of Rotating Generalized ECKS Black hole}

In this section, we employ the Hamilton--Jacobi formalism to study the null geodesic equations in the rotating generalized ECKS black hole spacetime. Our aim is to calculate the celestial coordinates parametrized with the radius of the unstable null orbits. Then, we  shall obtain the shadow of rotating generalized ECKS black hole. 
To this end, we first describe the motion of the particle on the rotating generalized ECKS black hole by the following Lagrangian:
\begin{equation}
\mathcal{L}=\frac{1}{2}g_{\nu \sigma }\dot{x}^{\nu }\dot{x}^{\sigma },
\label{r}
\end{equation}%
where $\dot{x}^{\nu }=u^{\nu }=dx^{\nu }/d\lambda $; let us recall that $u^{\nu }$ is 
four velocity of particle with the affine parameter $\lambda $. Since the conjugate momenta \ $p_{t}$ and $p_{\phi }$ are conserved because of the symmetry of black hole, the metric does not depend on the variables $t$ and $\phi $.  Afterwards, we can write energy $E$ and angular momentum $L$:
\begin{equation}
E=p_{t}=\frac{\partial \mathcal{L}}{\partial \dot{t}}=g_{\phi t}\dot{\phi}%
+g_{tt}\dot{t},\quad L=-p_{\phi }=-\frac{\partial \mathcal{L}}{\partial \dot{%
\phi}}=-g_{\phi \phi }\dot{\phi}-g_{\phi t}\dot{t}.  \label{E}
\end{equation}%
Then, one can get 
\begin{eqnarray}
&&\Sigma \dot{t}=-a(aE\sin ^{2}\theta -L)+\frac{(r^{2}+a^{2})P(r)}{\Delta (r)%
}, \\
&&\Sigma \dot{\phi}=-\left( aE-\frac{L}{\sin ^{2}\theta }\right) +\frac{aP(r)%
}{\Delta (r)},
\end{eqnarray}%
in which $P(r)\equiv E(r^{2}+a^{2})-aL$. To find the geodesics equations, we use the Hamilton-Jacobi (HJ) equation: 
\begin{equation}
\frac{\partial S}{\partial \lambda }=\frac{1}{2}g^{\nu \sigma }\frac{%
\partial S}{\partial x^{\nu }}\frac{\partial S}{\partial x^{\sigma }}.
\end{equation}
One can define the following ansatz as follows
\begin{equation}
S=\frac{1}{2}\mu ^{2}\lambda -Et+L\phi +S_{r}(r)+S_{\theta }(\theta ).
\end{equation}
Note that $\mu$ is proportional to the rest mass of the particle. For the stationary black hole spacetime, the HJ equation becomes: 
\begin{equation}
\frac{1}{2}g^{tt}\frac{\partial S}{\partial x^{t}}\frac{\partial S}{\partial
x^{t}}+g^{\phi t}\frac{\partial S}{\partial x^{t}}\frac{\partial S}{\partial
x^{\phi }}+\frac{1}{2}g^{rr}\frac{\partial S}{\partial x^{r}}\frac{\partial S%
}{\partial x^{r}}+\frac{1}{2}g^{\theta \theta }\frac{\partial S}{\partial
x^{\theta }}\frac{\partial S}{\partial x^{\theta }}+\frac{1}{2}g^{\phi \phi }%
\frac{\partial S}{\partial x^{\phi }}\frac{\partial S}{\partial x^{\phi }}=-%
\frac{\partial S}{\partial \lambda }.
\end{equation}
The solutions for the $S_r$ and $S_\theta$, respectively, are given by \cite{Johannsen:2015qca}:

\begin{eqnarray}
&&\Sigma \frac{\partial S_{r}}{\partial r}=\pm \sqrt{R(r)},
\label{eq:Theta_motion} \\
&&\Sigma \frac{\partial S_{\theta }}{\partial \theta }=\pm \sqrt{\Theta
(\theta )},
\end{eqnarray}%
with
\begin{eqnarray}
&&R(r)\equiv P(r)^{2}-\Delta (r)\left[ (L-aE)^{2}+\mathcal{Q}\right] , \\
&&\Theta (\theta )\equiv \mathcal{Q}+\cos ^{2}\theta \left( a^{2}E^{2}-\frac{%
L^{2}}{\sin ^{2}\theta }\right) .
\end{eqnarray}%
Here, $R$ and $\Theta$ are effective potentials for moving particle in radial $r$ and angular $\theta$ directions, respectively. Note that Carter constant  can be calculated as follows: $\mathcal{Q}\equiv 
\mathcal{K}-(L-aE)^{2}$ where $\mathcal{K}$ is a constant of motion. $R(r)$ and $\Theta (\theta )$ should be
positive for the photon motion. We can introduce the impact parameters $\eta $ and $\xi $:
\begin{equation}
\xi \equiv \frac{L}{E},\quad \quad \eta \equiv \frac{\mathcal{Q}}{E^{2}},
\end{equation}%
where $E$ is the energy and $L$  stands for the angular momentum. Eq.~(\ref{r}) can be rewritten in terms of dimensionless quantities $\eta $ and $\xi $ for the photon case: 
\begin{equation}
R(r)=\frac{1}{{E}^{2}}\left[ (r^{2}+a^{2})-a{\xi }\right] ^{2}-\Delta \left[
(a-{\xi })^{2}+{\eta }\right] .
\end{equation}
Afterwards, the equation of $S_r$ is obtained as follows \cite{Shaikh:2019fpu}:
\begin{equation}
( \frac{\partial S_r}{\partial r})^2 + V_{eff} = 0,
\end{equation}
where the effective potential $V_{eff}$:
\begin{equation}
V_{eff}=\frac{1}{{{\Sigma }^{2}}}\left[ (r^{2}+a^{2})-a{\xi }\right]
^{2}-\Delta \left[ (a-{\xi })^{2}+{\eta }\right] .  \label{vef}
\end{equation}%
To find the unstable circular orbits, we maximize the effective potential:
\begin{equation}
V_{eff}=\left. \frac{\partial V_{eff}}{\partial r}\right\vert
_{r=r_{0}}=0\;\;\;\;\mbox{or}\;\;\;R=\left. \frac{\partial R}{\partial r}%
\right\vert _{r=r_{0}}=0,  \label{vr}
\end{equation}%
in which $r=r_{0}$ is the radius of the unstable circular null orbit. It is noted that we locate the photons and the observer at the infinity ($\mu =0$) and assume that
photons come near to the equatorial plane $(\theta =\frac{\pi }{2})$. Then, we solve Eq. (\ref{vr}) to find the celestial coordinates:
\begin{eqnarray}
\xi  &=&{\frac{r^{2}-r\Delta -a^{2}}{a(r-1)}},  \label{17} \\
\eta  &=&{\frac{r^{3}[4\Delta -r(r-1)^{2}]}{a^{2}(r-1)^{2}}}.  \label{17aa}
\end{eqnarray}%
Here $r$ corresponds to the radius of the unstable null orbits. The apparent shape of the shadow cast is found by using the celestial coordinates \cite{Hioki:2009na}:
\begin{eqnarray}
Y&=&\lim\limits_{r_{0}\to \infty}\left(-r_{0}^{2}\sin\theta_{0}\frac{d\phi}{dr}\bigg|_{(r_{0,\theta_{0}})}\right), \\
X&=&\lim\limits_{r_{0}\to\infty}\left(r_{0}^{2}\frac{d\theta}{dr}\bigg|_{(r_{0},\theta_{0})}\right),
\end{eqnarray}
in which $(r_{0},\theta_{0})$ is for the coordinates of the observer. Hence, the
limiting the celestial coordinates become
\begin{eqnarray}
Y&=&-\frac{\xi}{\sin\theta_{0}}\label{alpha} ,\\
X&=&\pm\sqrt{\eta+a^{2}\cos\theta^{2}_{0}-\chi^{2}\cot^{2}\theta_{0}}\label{beta} \ ,
\end{eqnarray}
where the shadow corresponds to the parametric curve of $Y$ and $X$ in which $r$ stands as a parameter. Note that there is a special case in which the observer is on the
equatorial plane of the black hole with the inclination
angle $\theta_0= \pi/2$ \cite{young}. Hence, we have
\begin{equation}
\begin{aligned}
&Y=-\xi\\
&X=\pm \sqrt{\eta}
\end{aligned}
\end{equation}
Then the radius of the shadow can be calculated as follows:
\begin{equation} Y^{2}+X^{2}=\xi^{2}+\eta=R_s^2 \end{equation}

Hence \begin{equation}
  \begin{array}{c}
\xi=\frac{2 r_{0}\left(2 \gamma m r_{0}-q^2\right)-\left(r_{0}+\gamma m\right)\left(r_{0}^{2}+a^{2}\right)}{a\left(r_{0}-\gamma m\right)} \\
\eta=\frac{4 a^{2} r_{0}^{2}\left(\gamma m r_{0}-q^2\right)-r_{0}^{2}\left[r_{0}\left(r_{0}-3 \gamma m\right)+2 q^2\right]^{2}}{a^{2}\left(r_{0}-\gamma m\right)^{2}}
\end{array}
\end{equation}

 Shadows of the rotating generalized ECKS black hole having different values of spin $a$ and
parameter $\gamma$  are depicted in Figs. (3-10). The region bounded by each curve corresponds to the
black hole's shadow where the observers are located at spatial infinity and in the equatorial
plane ($i = \protect\pi/2$). Region in angular momentum space is occupied by the plunge orbits for particles in parabolic orbits, or photons,
incident upon a black hole from infinity.
Left/right side of the figures correspond to the prograde and retrograde circular photon orbits, respectively.

It is worth noting that since the rotating ECKS black hole under consideration is a generalization of the Kerr-Newman black hole. Once the rotation ceases, ECKS black hole reduces to the Reissner-Nordstr\"{o}m black hole for $\gamma=1$ and thus to the Schwarzschild black hole ($q=0$). It turns out that the shadow of a ECKS black hole is a zone covered by a deformed circle. It can be deduced from Figs. (3-10) that the shadow of a black hole is affected by the parameters $a$ and $\gamma$. Indeed, for a given $q$, the size of a shadow decreases as the parameter $a$ increases and the shadow becomes more distorted as we increase the value of parameter $\gamma$.

\begin{figure}[htbp]
\centering
\begin{minipage}[t]{0.44 \textwidth}

    \includegraphics[scale=0.70]{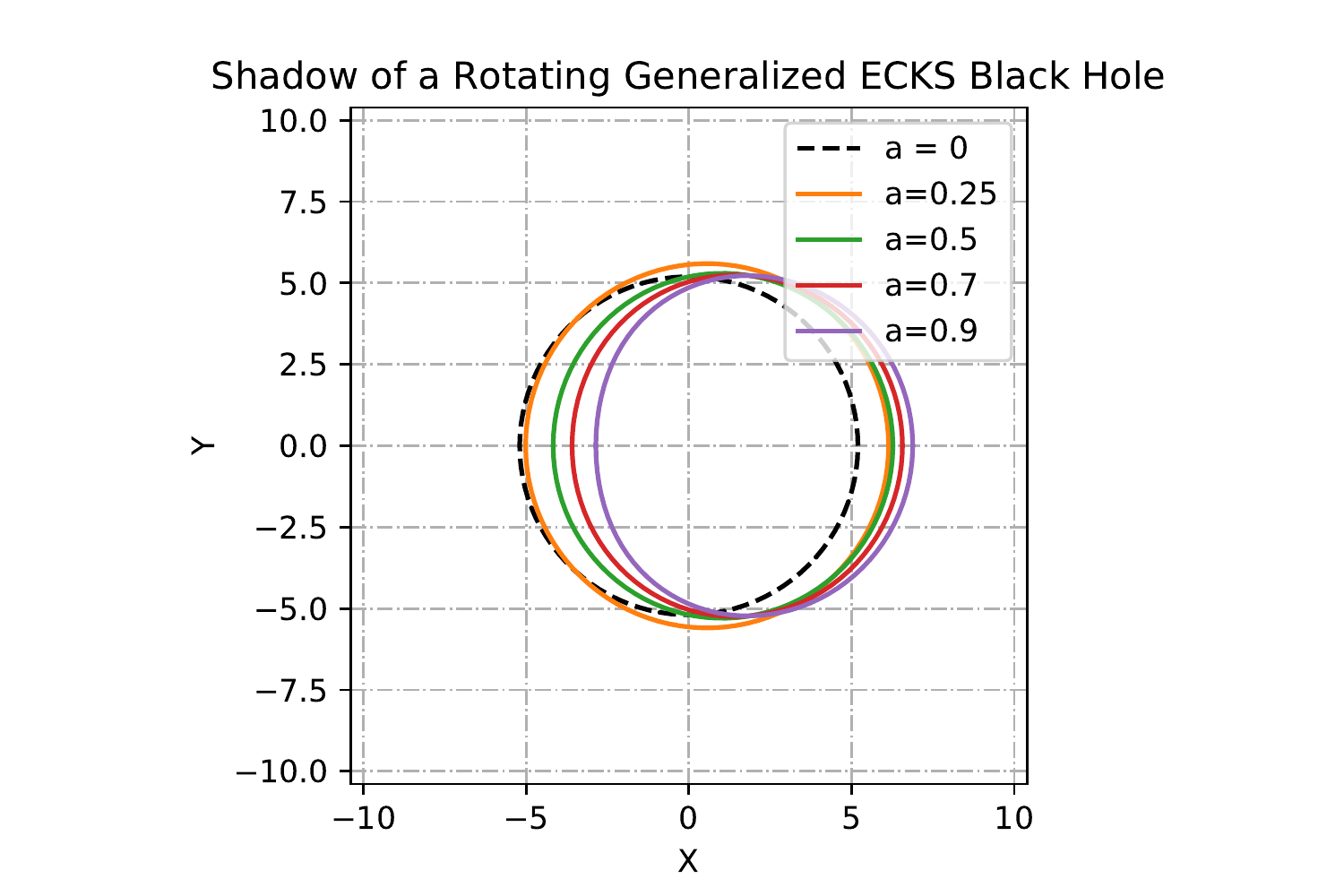}
    \caption{ $m=\gamma=1$ and the plots are for $q=0.1$. The dashed line is for the Schwarzschild black hole. }
       \label{fig:1}
       \end{minipage}
\hfill
\begin{minipage}[t]{0.47\textwidth}
    \includegraphics[scale=0.70]{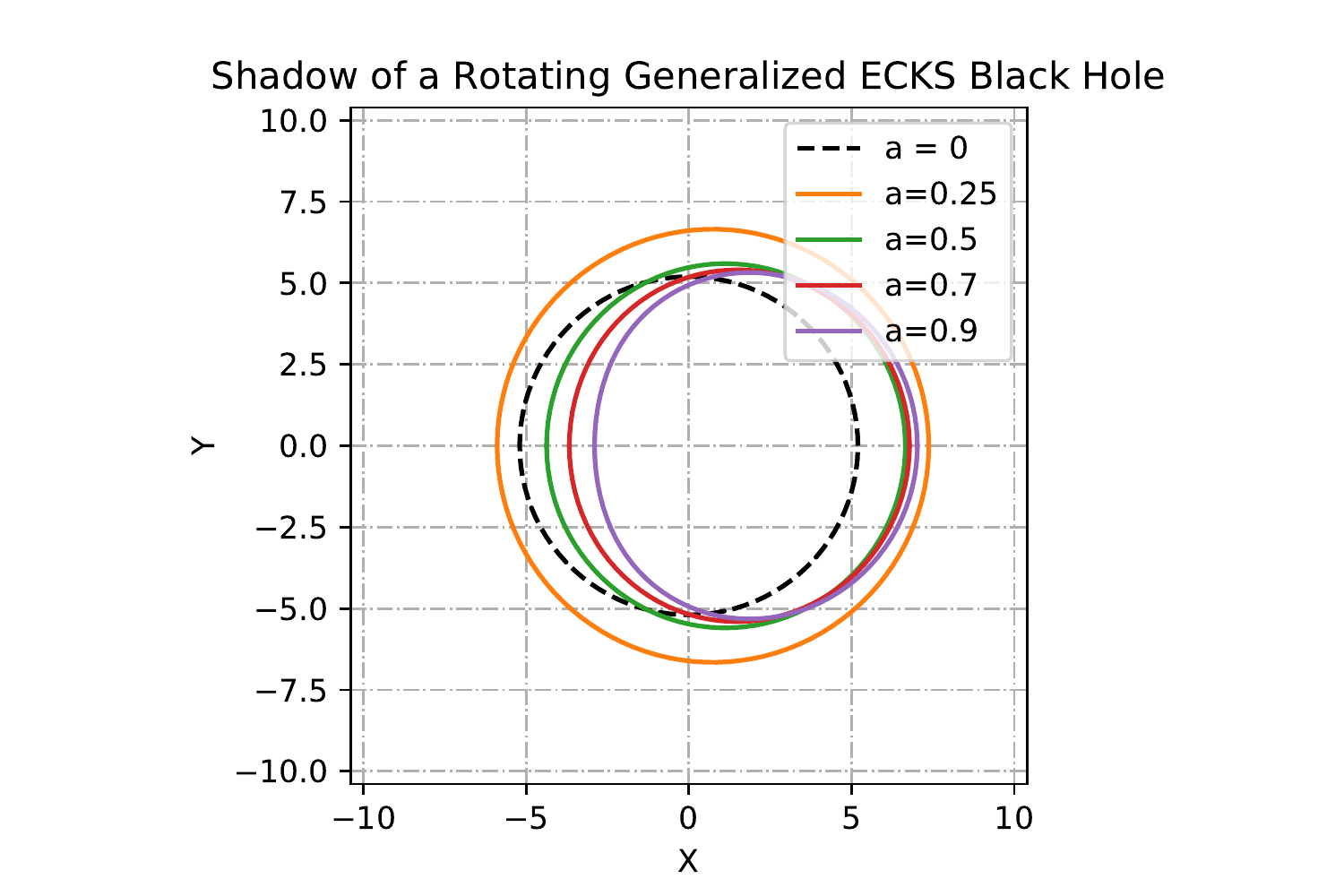}
    \caption{ $m=\gamma=1$ and the plots are for  $q=0.2$. The dashed line is for the Schwarzschild black hole. }
       \label{fig:2}
       \end{minipage}
\end{figure}
%%%%2
\begin{figure}[htbp]
\centering
\begin{minipage}[t]{0.44 \textwidth}
    \includegraphics[scale=0.70]{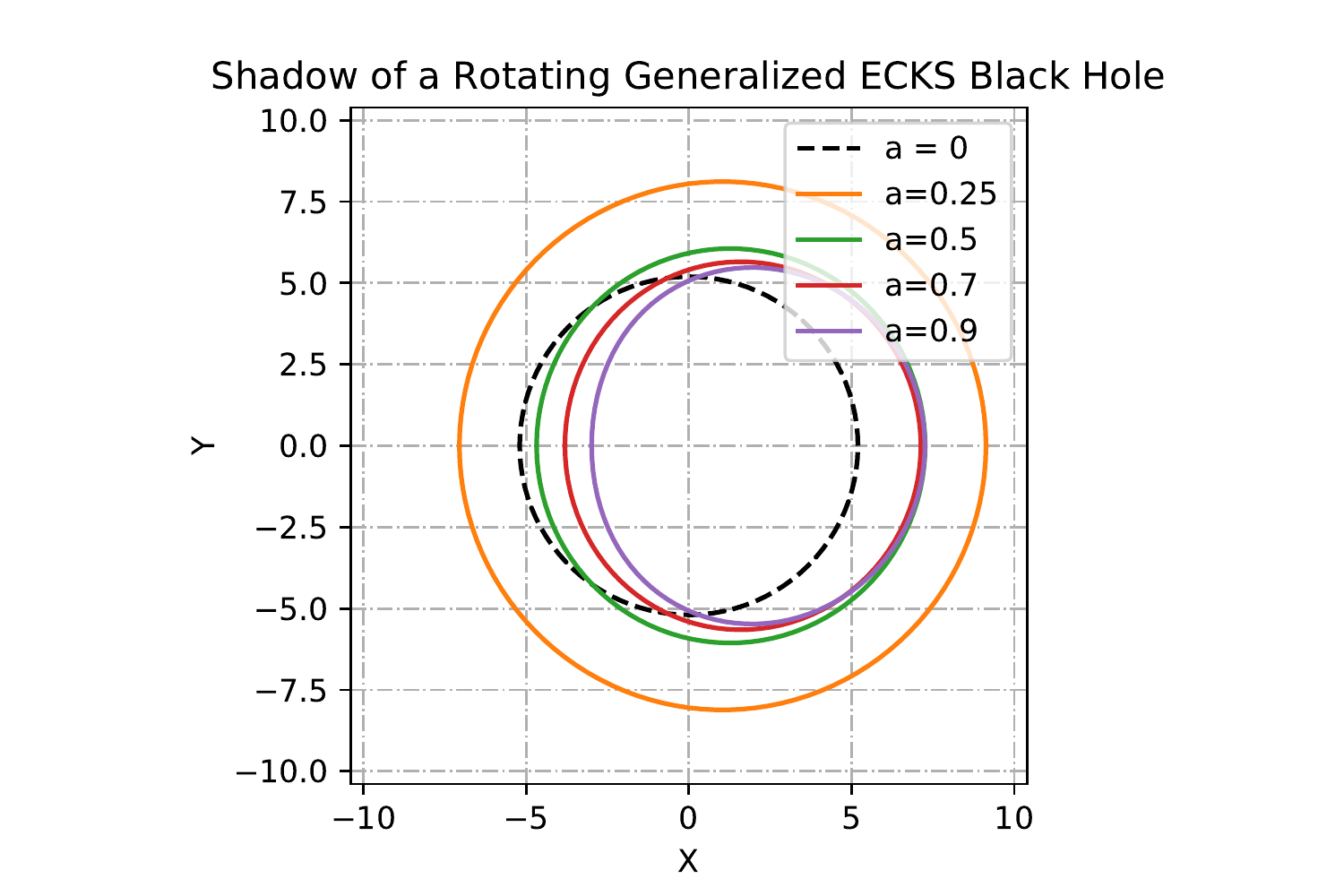}
\caption{ $m=\gamma=1$ and the plots are for $q=0.3$. The dashed line is for the Schwarzschild black hole. }
       \label{fig:3}
       \end{minipage}
\hfill
    \begin{minipage}[t]{0.47\textwidth}
    \includegraphics[scale=0.70]{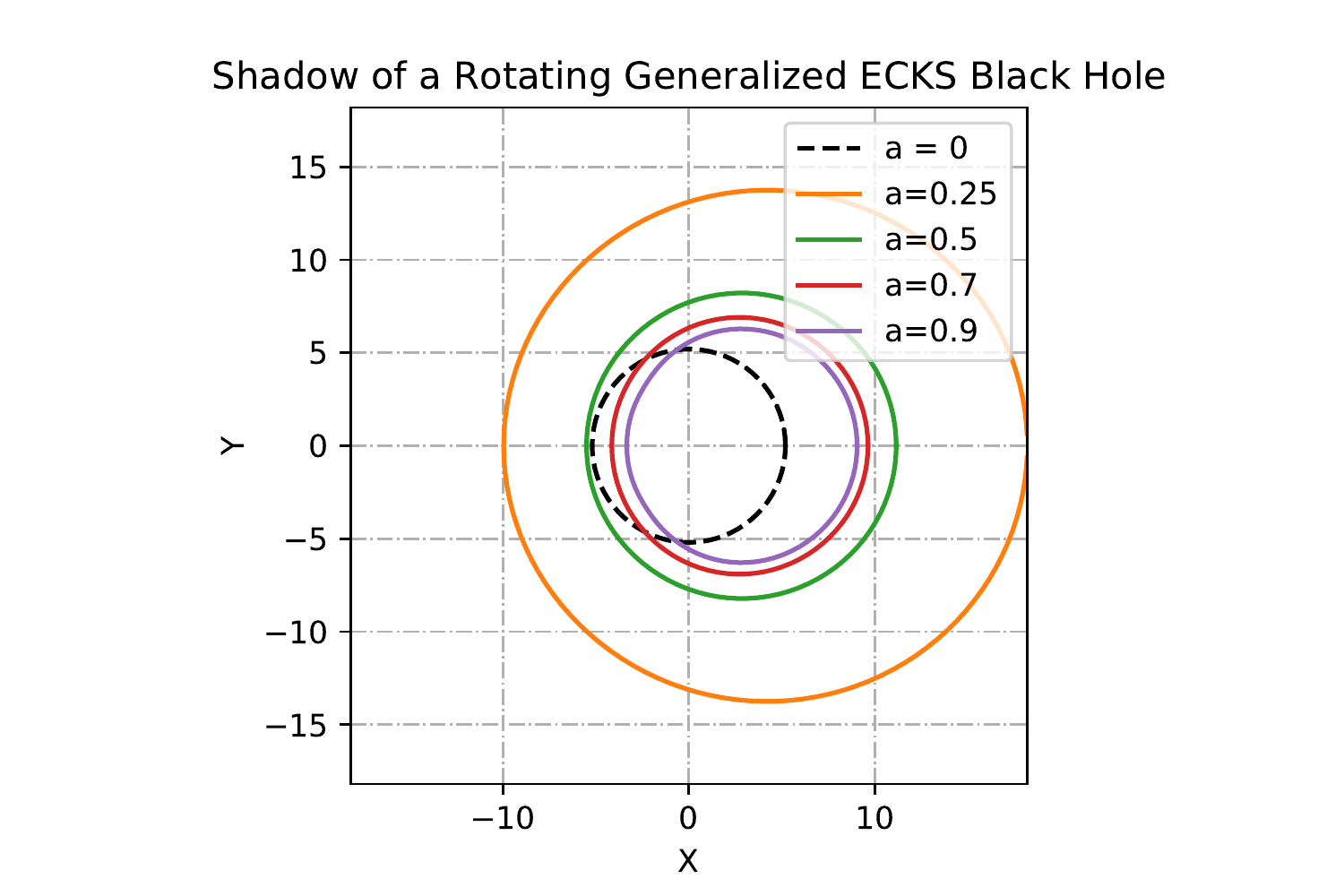}
    \caption{ $m=1$ and $q=0.1$, the plots are for $\gamma=0.94$. The dashed line is for the Schwarzschild black hole. }
       \label{fig:4}
              \end{minipage}
\end{figure}
%%%%3
\begin{figure}[htbp]
\centering
\begin{minipage}[t]{0.44 \textwidth}
    \includegraphics[scale=0.70]{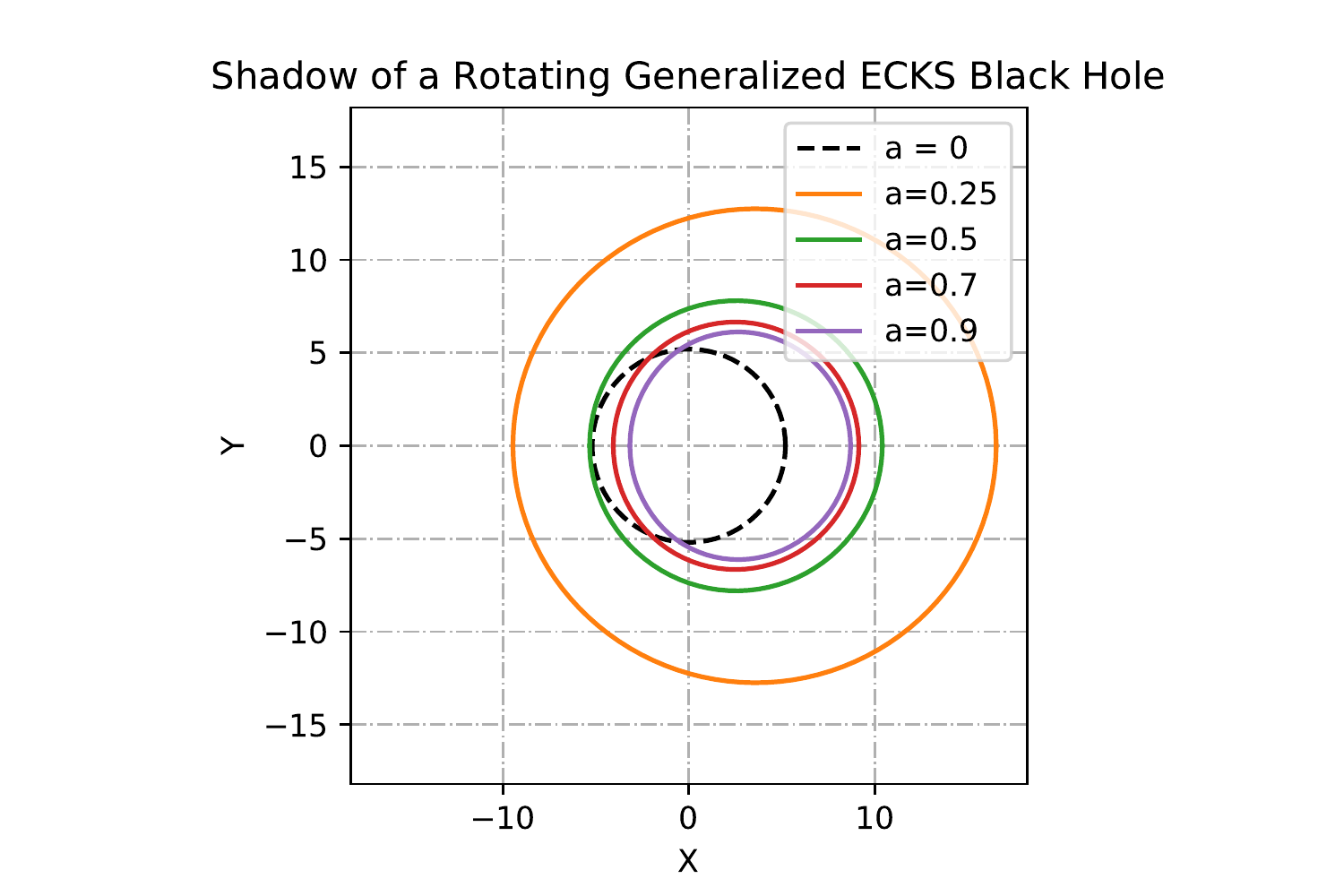}
    \caption{ $m=1$ and $q=0.1$, the plots are for  $\gamma=0.95$. The dashed line is for the Schwarzschild black hole. }
       \label{fig:5}
   \end{minipage}
\hfill
\begin{minipage}[t]{0.47\textwidth}
    \includegraphics[scale=0.70]{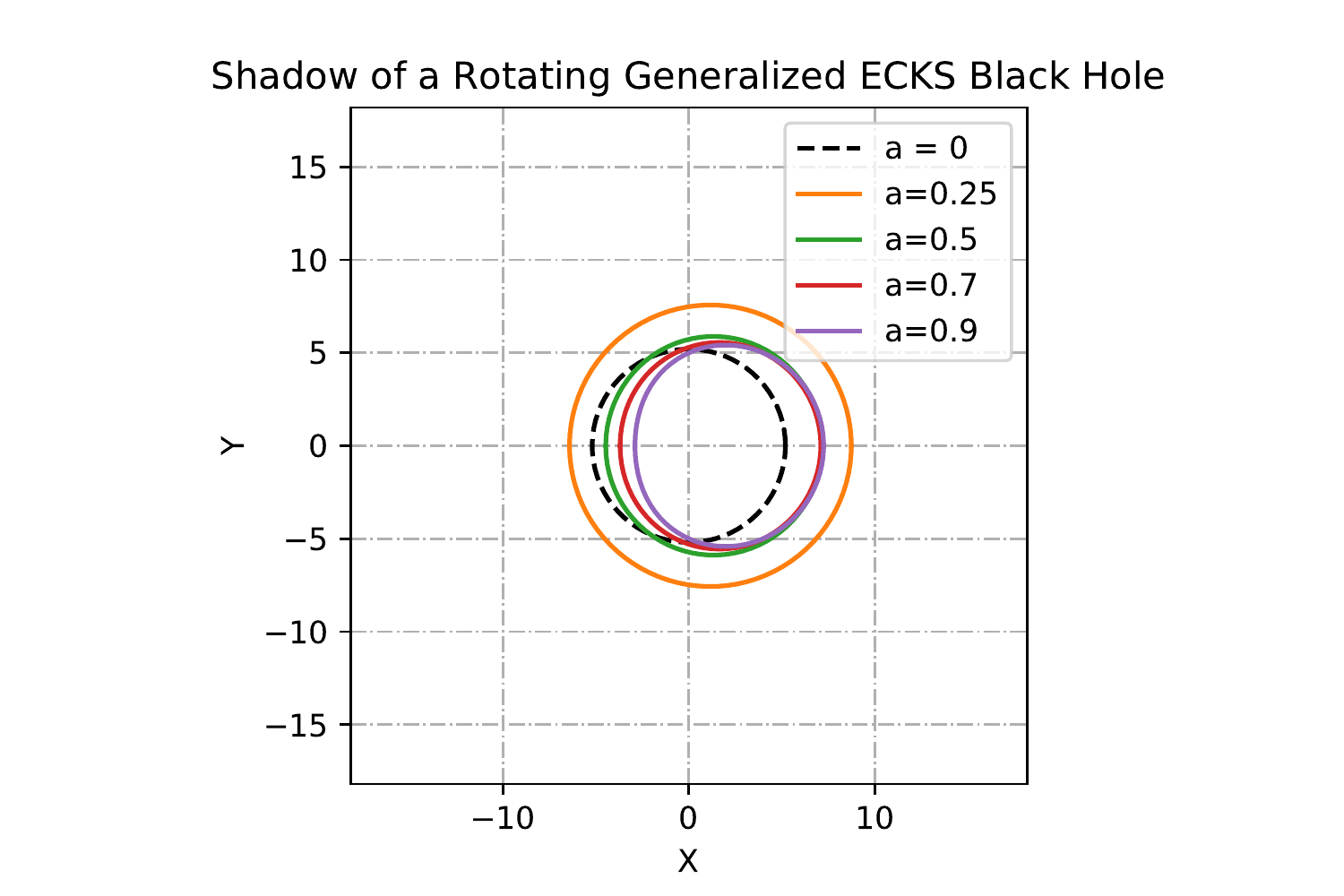}
\caption{ $m=1$ and $q=0.1$, the plots are for $\gamma=0.99$. The dashed line is for the Schwarzschild black hole. }
       \label{fig:6}
  \end{minipage}
\end{figure}
%%%%4

\begin{figure}[htbp]
\centering
\begin{minipage}[t]{0.44 \textwidth}
    \includegraphics[scale=0.70]{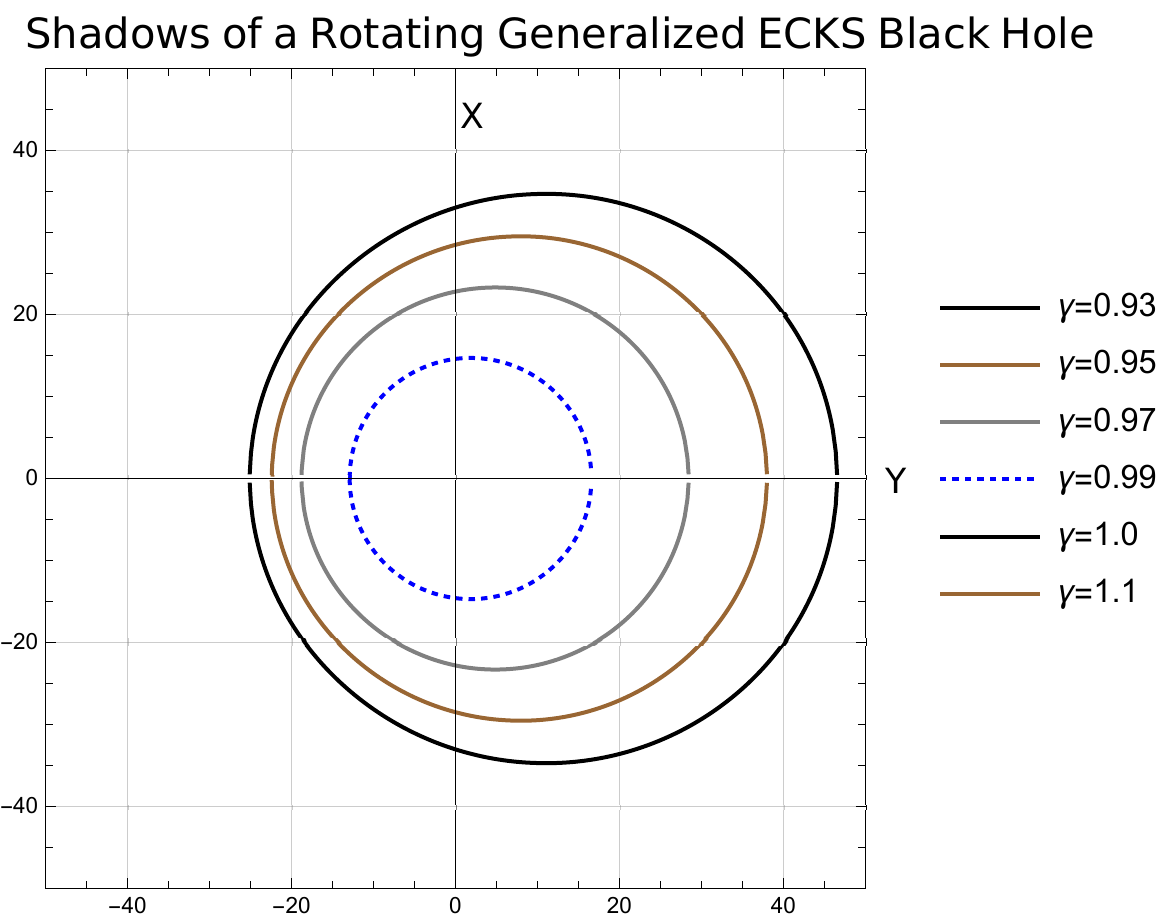}
\caption{ $m=1$ and $a=q=0.1$, the plots are for different values of $\gamma$.}
       \label{fig:6}
   \end{minipage}
\hfill
\begin{minipage}[t]{0.47\textwidth}
    \includegraphics[scale=0.70]{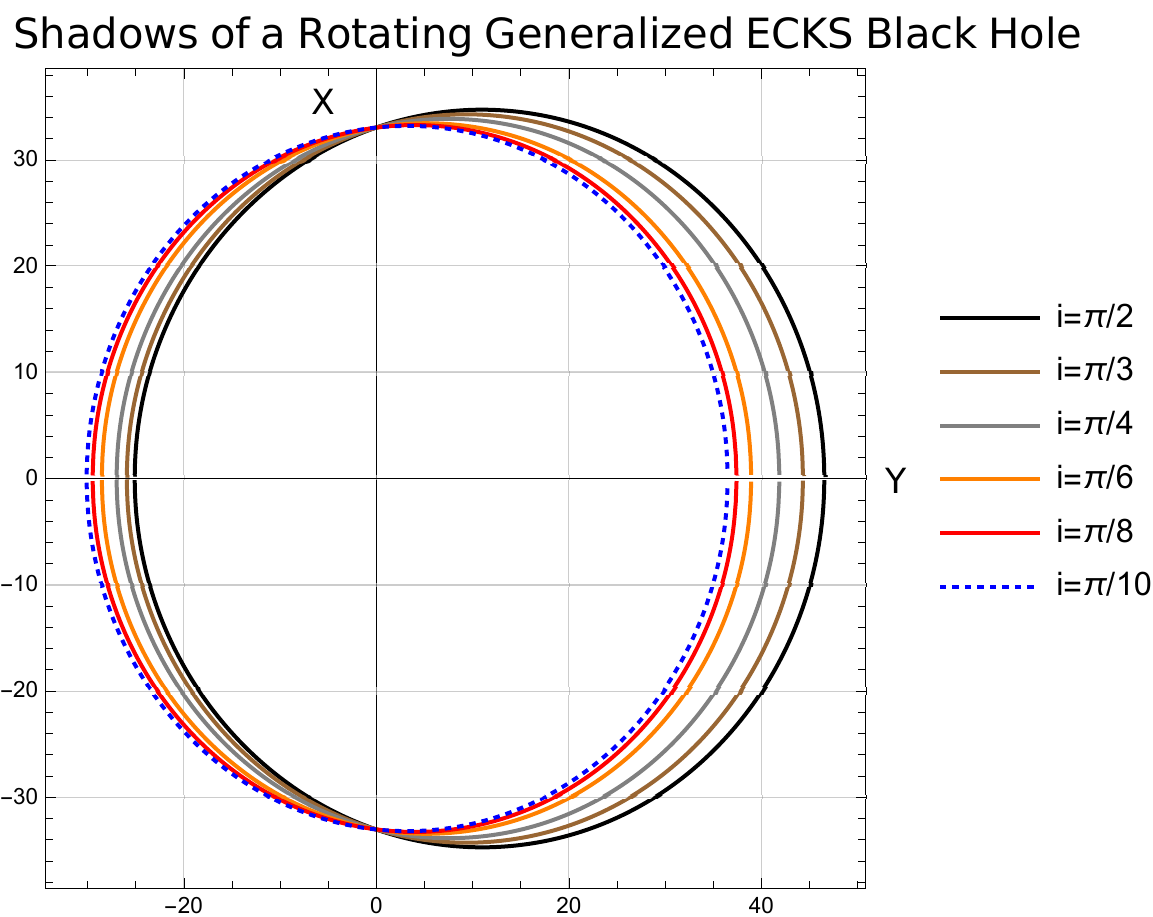}
\caption{Photon rings are shown at inclination angles $i$ for $m=1$, $a=q=0.1$, and $\gamma=0.93$. }
       \label{fig:6}
\end{minipage}
\end{figure}

\section{Conclusions}

In this paper, we have made a detailed analysis of the gravitational lensing problem of four dimensional ECKS black hole in the weak field approximation. For this purpose, we have first considered the static ECKS black hole. After employing the Gauss-Bonnet theorem and a straight line approach, we have obtained the light deflection angle by the static ECKS black hole at the leading order terms. Furthermore, we have also computed the deflection angle of light from the static ECKS black hole, which is in a plasma medium. Then, we have extended our study to the rotating version of this black hole. To derive the stationary ECKS black hole solution, we have used the Newman-Janis algorithm without considering the complexification. Then, we have thoroughly discussed  the gravitational lensing in the rotating ECKS black hole geometry.

For both cases (static and stationary), we have shown that the ECKS parameter $\gamma$ plays an important role on the path of the photons moving in the curved spacetime of the ECKS black hole. In particular, it is seen that the weak deflection is increased with the increase of ECKS parameter $\gamma$; the latter remark may shed light on the presence of the ECKS black holes in future cosmological observations. Another remarkable point is that as $\omega_{e} / \omega_{\infty} \rightarrow 0$, the plasma effect on the deflection angle is vanished. It is also worth noting that the deflection angle obtained with the Gauss-Bonnet theorem has been computed by taking the integral over the particular domain, which is outside the impact parameter. Thus, our gravitational lensing computations include the global effects. Shadows of the rotating ECKS black hole with different values of spin $a$ and
parameter $\gamma$ have been depicted in Figs. (3-10).  

Finally, our findings have the potential to indirectly indicate the presence of the torsion that may be the source of dark energy, a mysterious type of energy that permeates all of cosmos and increases the Universe's rate of expansion \cite{Lepe}. Therefore, due to its impact on the gravitational lensing, indirect proof of the torsion will provide a sound foundation for the scenario in which each black hole's interior is a new Universe. 

\section*{Acknowledgements}

The authors are grateful to the Editor and anonymous Referees for their valuable comments and suggestions to improve the paper.


\begin{thebibliography}{99}
\bibliographystyle{apsrev-nourl}

\bibitem{Abbott:2016blz}
Phys. Rev. Lett. \textbf{116} no.6, 061102 (2016).

\bibitem{Akiyama:2019cqa}
K.~Akiyama \textit{et al.} [Event Horizon Telescope],
Astrophys. J. \textbf{875} no.1, L1 (2019).


\bibitem{cartan}
E. Cartan, C. R. Acad. Sci. (Paris) {\bf 174}, 593 (1922); E. Cartan, Ann. Ec. Norm. Sup. {\bf 40}, 325 (1923); E. Cartan,
Ann. Ec. Norm. Sup. {\bf 41}, 1 (1924); E. Cartan, Ann. Ec. Norm. Sup. {\bf 42}, 17 (1925).

\bibitem{cartan2}
S.~M.~Khanapurkar and P.~Singh,
%``The Einstein-Cartan-Dirac theory,''
arXiv:1803.10621 [gr-qc].


\bibitem{cartan3}
F.~W.~Hehl, J.~McCrea, E.~W.~Mielke and Y.~Ne'eman,
%``Metric affine gauge theory of gravity: Field equations, Noether identities, world spinors, and breaking of dilation invariance,''
Phys. Rept. \textbf{258}, 1-171 (1995).

\bibitem{cartan4}
H.~Arcos and J.~Pereira,
%``Torsion gravity: A Reappraisal,''
Int. J. Mod. Phys. D \textbf{13}, 2193-2240  (2004).

\bibitem{cartan5}
P.~Baekler and F.~W.~Hehl,
%``Beyond Einstein-Cartan gravity: Quadratic torsion and curvature invariants with even and odd parity including all boundary terms,''
Class. Quant. Grav. \textbf{28}, 215017 (2011).

\bibitem{Tecchiolli:2019hfe}
M.~Tecchiolli,
%``On the Mathematics of Coframe Formalism and Einstein–Cartan Theory—A Brief Review,''
Universe \textbf{5}, no.10, 206 (2019).

\bibitem{Sciama}T. W. B. Kibble, J. Math. Phys. 2, 212 (1961); D. W. Sciama, Rev. Mod. Phys. 36, 463 (1964). 

\bibitem{Poplawski:2012ab}
N.~J.~Poplawski,
%``Big bounce from spin and torsion,''
Gen. Rel. Grav. \textbf{44}, 1007-1014 (2012).


\bibitem{sciama1} E. Sezgin, P. V. Nieuwenhuizen, Phys. Rev. D, 21 3269 (1980).

\bibitem{sciama2}J. A. R. Cembranos and J. G. Valcarcel,J. Cosmol. Astropart. Phys. 01, 014 (2017).



\bibitem{Chen:2018szr} 
  S.~Chen, L.~Zhang and J.~Jing,
 % ``A new asymptotical flat and spherically symmetric solution in the generalized Einstein-Cartan-Kibble-Sciama gravity and gravitational lensing,''
  Eur.\ Phys.\ J.\ C {\bf 78}, no. 11, 981 (2018).

\bibitem{izr1}
P. Schneider, J. Ehlers, and E.E. Falco, \textit{Gravitational Lenses} (Springer-Verlag, Berlin, 1992).

\bibitem{izr2}
A.O. Petters, H. Levine, J. Wambsganss, \textit{Singularity Theory and Gravitational Lensing} (Birkhäuser, Basel, 2001).

\bibitem{izr3}
X. Lu and Y. Xie, Eur. Phys. J. C \textbf{79}, 1016 (2019).

\bibitem{izr4}
Event Horizon Telescope Collaboration, Astrophys. J. Lett. \textbf{875}, L1 (2019).

\bibitem{izr5}
Event Horizon Telescope Collaboration, Astrophys. J. Lett. \textbf{875}, L2 (2019).

\bibitem{izr6}
Event Horizon Telescope Collaboration, Astrophys. J. Lett. \textbf{875}, L3 (2019).

\bibitem{izr7}
Event Horizon Telescope Collaboration, Astrophys. J. Lett. \textbf{875}, L4 (2019).

\bibitem{izr8}
Event Horizon Telescope Collaboration, Astrophys. J. Lett. \textbf{875}, L5 (2019).

\bibitem{izr9}
Event Horizon Telescope Collaboration, Astrophys. J. Lett. \textbf{875}, L6 (2019).

\bibitem{izr10}
F.~Roelofs et al.,
%``Simulations of imaging the event horizon of Sagittarius A* from space,''
Astron. Astrophys. \textbf{625}, A124 (2019).

\bibitem{Blake:2020mzy}
C.~Blake, et. al,
%``Testing gravity using galaxy-galaxy lensing and clustering amplitudes in KiDS-1000, BOSS and 2dFLenS,''
arXiv:2005.14351 [astro-ph.CO].

\bibitem{Joachimi:2020abi}
B.~Joachimi, et. al,
%``KiDS-1000 Methodology: Modelling and inference for joint weak gravitational lensing and spectroscopic galaxy clustering analysis,''
arXiv:2007.01844 [astro-ph.CO].

\bibitem{Monteiro-Oliveira:2020yfy}
R.~Monteiro-Oliveira, et. al,
%``Revising the merger scenario of the galaxy cluster Abell 1644: a new gas poor structure discovered by weak gravitational lensing,''
Mon. Not. Roy. Astron. Soc. \textbf{495}, no.2, 2007-2021 (2020).

\bibitem{Yu:2020agu}
H.~Yu, P.~Zhang and F.~Y.~Wang,
%``Strong lensing as a giant telescope to localize the host galaxy of gravitational wave event,''
Mon. Not. Roy. Astron. Soc. \textbf{497}, no.1, 204-209 (2020).

\bibitem{Foxley-Marrable:2020ckf}
M.~Foxley-Marrable, et. al
%``Observing the earliest moments of supernovae using strong gravitational lenses,''
doi:10.1093/mnras/staa1289
ArXiv:2003.14340 [astro-ph.HE].

\bibitem{Tu:2019vcj}
Z.~L.~Tu, J.~Hu and F.~Y.~Wang,
%``Probing cosmic acceleration by strong gravitational lensing systems,''
Mon. Not. Roy. Astron. Soc. \textbf{484}, no.3, 4337-4346 (2019).

\bibitem{Qi:2019zdk}
J.~Z.~Qi and X.~Zhang,
%``A new cosmological probe using super-massive black hole shadows,''
Chin. Phys. C \textbf{44}, no.5, 055101 (2020).



\bibitem{edd} F. W. Dyson,A. S. Eddington, C. Davidson, Philosophical Transactions of the Royal Society A: Mathematical, Physical and Engineering Sciences. 220 (571–581): 291–333 (1920). 
%%

\bibitem{Bartelmann:1999yn} 
  M.~Bartelmann and P.~Schneider,
  %``Weak gravitational lensing,''
  Phys.\ Rept.\  {\bf 340}, 291 (2001).
  
  
  \bibitem{Bozza:2002zj}
V.~Bozza,
%``Gravitational lensing in the strong field limit,''
Phys. Rev. D \textbf{66}, 103001 (2002).

\bibitem{Tsukamoto:2012xs}
N.~Tsukamoto, T.~Harada and K.~Yajima,
%``Can we distinguish between black holes and wormholes by their Einstein ring systems?,''
Phys. Rev. D \textbf{86}, 104062 (2012).

\bibitem{Aazami:2011tw}
A.~B.~Aazami, C.~R.~Keeton and A.~Petters,
%``Lensing by Kerr Black Holes. II: Analytical Study of Quasi-Equatorial Lensing Observables,''
J. Math. Phys. \textbf{52}, 102501 (2011).

\bibitem{Virbhadra:2007kw}
K.~Virbhadra and C.~Keeton,
%``Time delay and magnification centroid due to gravitational lensing by black holes and naked singularities,''
Phys. Rev. D \textbf{77}, 124014 (2008).

\bibitem{Ishak:2010zh}
M.~Ishak and W.~Rindler,
%``The Relevance of the Cosmological Constant for Lensing,''
Gen. Rel. Grav. \textbf{42}, 2247-2268 (2010).

\bibitem{Keeton:2005jd}
C.~R.~Keeton and A.~Petters,
%``Formalism for testing theories of gravity using lensing by compact objects. I. Static, spherically symmetric case,''
Phys. Rev. D \textbf{72}, 104006 (2005).

\bibitem{Wei:2014dka}
S.~Wei, K.~Yang and Y.~Liu,
%``Black hole solution and strong gravitational lensing in Eddington-inspired Born–Infeld gravity,''
Eur. Phys. J. C \textbf{75}, 253 (2015).


\bibitem{Iyer:2006cn}
S.~V.~Iyer and A.~O.~Petters,
%``Light's bending angle due to black holes: From the photon sphere to infinity,''
Gen. Rel. Grav. \textbf{39}, 1563-1582 (2007).

\bibitem{Gibbons:2011rh}
G.~Gibbons and M.~Vyska,
%``The Application of Weierstrass elliptic functions to Schwarzschild Null Geodesics,''
Class. Quant. Grav. \textbf{29}, 065016 (2012).

\bibitem{Aliev:2009cg}
A.~N.~Aliev and P.~Talazan,
%``Gravitational Effects of Rotating Braneworld Black Holes,''
Phys. Rev. D \textbf{80}, 044023 (2009).

\bibitem{Liu:2010wh}
Y.~Liu, S.~Chen and J.~Jing,
%``Strong gravitational lensing in a squashed Kaluza-Klein black hole spacetime,''
Phys. Rev. D \textbf{81}, 124017 (2010).

\bibitem{Sahu:2012er}
S.~Sahu, M.~Patil, D.~Narasimha and P.~S.~Joshi,
%``Can strong gravitational lensing distinguish naked singularities from black holes?,''
Phys. Rev. D \textbf{86}, 063010 (2012).

\bibitem{Sereno:2003nd}
M.~Sereno,
%``Weak field limit of Reissner-Nordstrom black hole lensing,''
Phys. Rev. D \textbf{69}, 023002 (2004).

\bibitem{Iyer:2009wa}
S.~Iyer and E.~Hansen,
%``Light's Bending Angle in the Equatorial Plane of a Kerr Black Hole,''
Phys. Rev. D \textbf{80}, 124023 (2009).

\bibitem{Sotani:2015ewa}
H.~Sotani and U.~Miyamoto,
%``Strong gravitational lensing by an electrically charged black hole in Eddington-inspired Born-Infeld gravity,''
Phys. Rev. D \textbf{92}, no.4, 044052 (2015).

\bibitem{Chen:2011ef}
S.~Chen, Y.~Liu and J.~Jing,
%``Strong gravitational lensing in a squashed Kaluza-Klein Godel black hole,''
Phys. Rev. D \textbf{83}, 124019 (2011).

\bibitem{Majumdar:2004mz}
A.~S.~Majumdar and N.~Mukherjee,
%``Gravitational lensing in the weak field limit by a braneworld black hole,''
Mod. Phys. Lett. A \textbf{20}, 2487-2496 (2005).

\bibitem{Tsupko:2013cqa}
O.~Y.~Tsupko and G.~S.~Bisnovatyi-Kogan,
%``Gravitational lensing in plasma: Relativistic images at homogeneous plasma,''
Phys. Rev. D \textbf{87} no.12, 124009 (2013).

\bibitem{Tsukamoto:2016zdu}
N.~Tsukamoto and T.~Harada,
%``Light curves of light rays passing through a wormhole,''
Phys. Rev. D \textbf{95} no.2, 024030 (2017).

\bibitem{Fernando:2011ki}
S.~Fernando,
%``Null Geodesics of Charged Black Holes in String Theory,''
Phys. Rev. D \textbf{85}, 024033 (2012)

\bibitem{Shaikh:2017zfl}
R.~Shaikh and S.~Kar,
%``Gravitational lensing by scalar-tensor wormholes and the energy conditions,''
Phys. Rev. D \textbf{96} no.4, 044037 (2017).

\bibitem{Bisnovatyi-Kogan:2017kii}
G.~S.~Bisnovatyi-Kogan and O.~Y.~Tsupko,
%``Gravitational Lensing in Presence of Plasma: Strong Lens Systems, Black Hole Lensing and Shadow,''
Universe \textbf{3} no.3, 57 (2017).

\bibitem{Tsukamoto:2017edq}
N.~Tsukamoto,
%``Retrolensing by a wormhole at deflection angles π and 3π,''
Phys. Rev. D \textbf{95} no.8, 084021 (2017).
\bibitem{pp3 1} A. Abdujabbarov, B. Toshmatov, J. Schee, Z. Stuchlik, and
B. Ahmedov, Int. J. Mod. Phys. D 26, 1741011 (2017).

\bibitem{pp3 2} A. Abdujabbarov, M. Amir, B. Ahmedov, and S. G. Ghosh, Phys. Rev. D 93, 104004 (2016).

\bibitem{pp3 3} A. Abdujabbarov, B. Ahmedov, N. Dadhich, and F. Atamurotov, Phys. Rev. D 96, 084017 (2017).

\bibitem{pp3 4} A. Abdujabbarov, B. Juraev, B. Ahmedov, and Z. Stuchlik, Astrophys. Space Sci. 361, 226 (2016).

\bibitem{pp3 5} A. Abdujabbarov, B. Toshmatov, Z. Stuchlik, and B.
Ahmedov, Int. J. Mod. Phys. Conf. Ser. 26, 1750051 (2017).

\bibitem{pp3 6} B. Turimov, B. Ahmedov, A. Abdujabbarov, and C. Bambi, Int. J. Mod. Phys. D \textbf{28} no.16, 2040013 (2019).

\bibitem{pp3 7} H. Chakrabarty, A. B. Abdikamalov, A. A. Abdujabarov,
and C. Bambi, Phys. Rev. D \textbf{98} no.2, 024022 (2018).

\bibitem{pp3 8} F. Atamurotov, B. Ahmedov, and A. Abdujabbarov, Phys. Rev. D 92, 084005 (2015).


\bibitem{AzregAinou:2012xv}
M.~Azreg-Ainou,
%``Light paths of normal and phantom Einstein-Maxwell-dilaton black holes,''
Phys. Rev. D \textbf{87} no.2, 024012 (2013).

\bibitem{Wang:2016paq}
S.~Wang, S.~Chen and J.~Jing,
%``Strong gravitational lensing by a Konoplya-Zhidenko rotating non-Kerr compact object,''
JCAP \textbf{11}, 020 (2016).

\bibitem{Sharif:2015kna}
M.~Sharif and S.~Iftikhar,
%``Null Geodesics and Strong Field Gravitational Lensing of Black Hole with Global Monopole,''
Adv. High Energy Phys. \textbf{2015}, 854264 (2015).



%%%

\bibitem{Gibbons:2008rj} 
  G.~W.~Gibbons and M.~C.~Werner,
 % ``Applications of the Gauss-Bonnet theorem to gravitational lensing,''
  Class.\ Quant.\ Grav.\  {\bf 25}, 235009 (2008)
  
    \bibitem{Werner2012} M. C. Werner, Gen. Relativ. Gravit. {\bf 44}, 3047 (2012).
  

\bibitem{Crisnejo:2018uyn} 
  G.~Crisnejo and E.~Gallo,
 % ``Weak lensing in a plasma medium and gravitational deflection of massive particles using the Gauss-Bonnet theorem. A unified treatment,''
  Phys.\ Rev.\ D {\bf 97}, no. 12, 124016 (2018)


  
  \bibitem{Jusufi:string17} K. Jusufi, I. Sakalli, and A. \"{O}vg\"{u}n, Phys. Rev. D {\bf 96}, 024040 (2017).
  
  \bibitem{Jusufi&Ali:Teo}A. \"{O}vg\"{u}n, K. Jusufi, and I. Sakalli, Ann. Phys. (Amsterdam) {\bf 399}, 193 (2018). 
  
  \bibitem{Jusufi&Ali:string} K. Jusufi and A. \"{O}vg\"{u}n, Phys. Rev. D {\bf 97}, 064030 (2018).
  
  \bibitem{Jusufi:RB} A. \"{O}vg\"{u}n, K. Jusufi, and I. Sakalli, Phys. Rev. D {\bf 99}, 024042 (2019). 
  
  \bibitem{Jusufi:monopole} K. Jusufi, M. C. Werner, A. Banerjee, and A. \"{O}vg\"{u}n, Phys. Rev. D {\bf 95}, 104012 (2017).
  
  \bibitem{Ali:BML}  W. Javed, R. Babar, and A. \"{O}vg\"{u}n, Phys. Rev. D {\bf 99}, 084012 (2019).
  
  \bibitem{Ali:wormhole}K. Jusufi, A. \"{O}vg\"{u}n, J. Saavedra, Y. Vasquez, and P. A. Gonzalez, Phys. Rev. D {\bf 97}, 124024 (2018).
  
  \bibitem{Ali:strings} A. \"{O}vg\"{u}n, Phys. Rev. D {\bf 99}, 104075 (2019).

  \bibitem{Javed1}K. Jusufi and A. \"{O}vg\"{u}n, Phys. Rev. D {\bf 97}, 024042 (2018).
  
  \bibitem{Javed2} W. Javed, R. Babar, and A. \"{O}vg\"{u}n, Phys. Rev. D {\bf 100}, 104032 (2019).
  
  \bibitem{Javed3} W. Javed, J. Abbas, A. \"{O}vg\"{u}n, Eur. Phys. J. C {\bf 79}, 694 (2019).
  
  \bibitem{Javed4}  W. Javed, M. Bilal Khadim, J. Abbas, A. {\"O}vg{\"u}n, Eur. Phys. J. Plus 135, 314 (2020).


  \bibitem{Sakalli2017} I. Sakalli and A. \"{O}vg\"{u}n, Europhys. Lett. {\bf 118}, 60006 (2017).
  
  \bibitem{Goulart2018} P. Goulart, Classical Quantum Gravity {\bf 35}, 025012 (2018).
  
  \bibitem{Leon2019} K. de Leon and I. Vega, Phys. Rev. D {\bf 99}, 124007 (2019).
  
  \bibitem{LZ20201} Z. Li and T. Zhou, Phys. Rev. D {\bf 101}, 044043 (2020).
  
  \bibitem{zhu2019} T. Zhu, Q. Wu, M. Jamil, and K. Jusufi, Phys.\ Rev.\ D {\bf 100}, no. 4, 044055 (2019).
  
  \bibitem{444}  A.~\"{O}vg\"{u}n, I.~Sakalli, and J.~Saavedra, Annals Phys.\   411, 167978 (2019).
  
  \bibitem{445} A.~\"{O}vg\"{u}n, I.~Sakalli, and J.~Saavedra, JCAP 1810, 041 (2018).
  
  \bibitem{446}  K.~Jusufi, A.~\"{O}vg\"{u}n, A.~Banerjee and I.~Sakalli, Eur.\ Phys.\ J.\ Plus 134, no. 9, 428 (2019).
  
  \bibitem{447} 
  A.~\"{O}vg\"{u}n, G.~Gyulchev, and K.~Jusufi, Annals Phys.\  406, 152 (2019).

  \bibitem{448} 
  K.~Jusufi and A. {\"O}vg{\"u}n,
  %``Light Deflection by a Quantum Improved Kerr Black Hole Pierced by a Cosmic String,''
  Int.\ J.\ Geom.\ Meth.\ Mod.\ Phys.\ 16, no. 08, 1950116 (2019).
  
    \bibitem{449} 
  Y.~Kumaran and A.~\"{O}vg\"{u}n, Chin.\ Phys.\ C 44, 025101 (2020).
  
  
   \bibitem{450} 
  W.~Javed, j.~Abbas and A.~\"{O}vg\"{u}n, Phys.\ Rev.\ D  100, no. 4, 044052 (2019). 
  
  \bibitem{1765621}
W.~Javed, A.~Hazma and A.~\"{O}vg\"{u}n,
%``Effect of Non-Linear Electrodynamics on Weak Field Deflection Angle by Black Hole,''
Phys. Rev. D \textbf{101}, no.10, 103521 (2020).
  
  
  \bibitem{CGJ2019}  A.~\"{O}vg\"{u}n,  Universe  5, 115 (2019).
  
  \bibitem{Jusufi:mp}   A. \"{O}vg\"{u}n, Phys.\ Rev.\ D  98, 044033 (2018).
  
  \bibitem{LHZ2020} Z. Li, G. He, and T. Zhou, Phys. Rev. D {\bf 101}, 044001 (2018).
  
  \bibitem{ISOA2016} A. Ishihara, Y. Suzuki, T. Ono, T. Kitamura, and H. Asada, Phys. Rev. D {\bf 94}, 084015 (2016).
  
  \bibitem{IOA2017} A. Ishihara, Y. Suzuki, T. Ono, and H. Asada, Phys. Rev. D {\bf 95}, 044017 (2017).
  
  \bibitem{OIA2017} T. Ono, A. Ishihara, and H. Asada, Phys. Rev. D {\bf 96}, 104037 (2017).
  
  \bibitem{OIA2018} T. Ono, A. Ishihara, and H. Asada, Phys. Rev. D {\bf 98}, 044047 (2018).
  
  \bibitem{OIA2019} T. Ono, A. Ishihara, and H. Asada, Phys. Rev. D {\bf 99}, 124030 (2019).
  
  \bibitem{OA2019} T. Ono and H. Asada, Universe \textbf{5}, no.11, 218 (2019).
  
\bibitem{Arakida2018} H. Arakida, Gen. Relativ. Gravit. {\bf 50}, 48 (2018).

  \bibitem{LA2020} Z. Li and A. \"{O}vg\"{u}n, Phys. Rev. D {\bf 101}, 024040 (2020).
  
  \bibitem{LJ2020} Z. Li and J. Jia, Eur. Phys. J. C {\bf 80}, 157 (2020).
  
  \bibitem{LZ20202} Z. Li and T. Zhou, arXiv:2001.01642.



  \bibitem{Takizawa2020} K. Takizawa, T. Ono, and H. Asada, Phys. Rev. D \textbf{101}, no.10, 104032 (2020).

  \bibitem{Gibbons2016} G. W. Gibbons, Classical Quantum Gravity {\bf 33}, 025004 (2016).
 
 \bibitem{Islam:2020xmy}
S.~U.~Islam, R.~Kumar and S.~G.~Ghosh,
%``Gravitational lensing by black holes in $4D$ Einstein-Gauss-Bonnet gravity,''
arXiv:2004.01038 [gr-qc].

\bibitem{Pantig:2020odu}
R.~C.~Pantig and E.~T.~Rodulfo,
%``Weak lensing of a dirty black hole,''
Chin. J. Phys. \textbf{66}, 691-702 (2020).

\bibitem{Takizawa:2020egm}
K.~Takizawa, T.~Ono and H.~Asada,
%``Gravitational deflection angle of light: Definition by an observer and its application to an asymptotically nonflat spacetime,''
Phys. Rev. D \textbf{101}, no.10, 104032 (2020).

\bibitem{Kumar:2020hgm}
R.~Kumar, S.~G.~Ghosh and A.~Wang,
%``Light Deflection and Shadow Cast by Rotating Kalb-Ramond Black Holes,''
Phys. Rev. D \textbf{101}, no.10, 104001 (2020).

\bibitem{Tsukamoto:2020uay}
N.~Tsukamoto,
%``Non-logarithmic divergence of a deflection angle by a marginally unstable photon sphere of the Damour-Solodukhin wormhole in a strong deflection limit,''
Phys. Rev. D \textbf{101}, no.10, 104021 (2020).

\bibitem{Crisnejo:2019xtp}
G.~Crisnejo, E.~Gallo and J.~R.~Villanueva,
%``Gravitational lensing in dispersive media and deflection angle of charged massive particles in terms of curvature scalars and energy-momentum tensor,''
Phys. Rev. D \textbf{100} no.4, 044006, (2019).


  %SHADOW
  
  %
  
\bibitem{synge} J. L. Synge, Mon. Not. R. Astron. Soc. \textbf{131}, 463 (1966).
  
  \bibitem{luminet} J. P. Luminet, Astronomy and Astrophysics \textbf{75}, 228 (1979).

\bibitem{cunha} P. V. P. Cunha, C. A. R. Herdeiro, E. Radu, and H. F.
Runarsson, Phys. Rev. Lett. \textbf{115}, 211102 (2015).

\bibitem{Cunha:2018acu}  P.~V.~P.~Cunha and C.~A.~R.~Herdeiro,  
%``Shadows and strong gravitational lensing: a brief review,''
Gen.\ Rel.\ Grav.\ \textbf{50}, no. 4, 42 (2018).


\bibitem{Falcke:1999pj}  H.~Falcke, F.~Melia and E.~Agol,  
%``Viewing the shadow of the black hole at the galactic center,''
Astrophys.\ J.\ \textbf{528}, L13 (2000).

\bibitem{Tremblay:2016ijg}  G.~R.~Tremblay \textit{et al.},  
%``Cold, clumpy accretion onto an active supermassive black hole,''
Nature \textbf{534}, 218 (2016).

%\cite{Shen:2005cw}

\bibitem{Shen:2005cw}  Z.~Q.~Shen, K.~Y.~Lo, M.-C.~Liang, P.~T.~P.~Ho and
J.-H.~Zhao,  
%``A size of ~1 au for the radio source sgr a* at the centre of the milky way,''
Nature \textbf{438}, 62 (2005).

%\cite{Huang:2007us}

\bibitem{Huang:2007us}  L.~Huang, M.~Cai, Z.~Q.~Shen and F.~Yuan,  
%``Black Hole Shadow Image and Visibility Analysis of Sagittarius A*,''
Mon.\ Not.\ Roy.\ Astron.\ Soc.\ \textbf{379}, 833 (2007).

%\cite{Johannsen:2015mdd}

\bibitem{Johannsen:2015mdd}  T.~Johannsen,  
%``Sgr A* and General Relativity,''
Class.\ Quant.\ Grav.\ \textbf{33}, no. 11, 113001 (2016).


\bibitem{Hioki:2009na}
K.~Hioki and K.~Maeda,
%``Measurement of the Kerr Spin Parameter by Observation of a Compact Object's Shadow,''
Phys. Rev. D \textbf{80}, 024042 (2009).
%\cite{Cunha:2018gql}

\bibitem{Cunha:2018gql}  P.~V.~P.~Cunha, C.~A.~R.~Herdeiro and
M.~J.~Rodriguez,  
%``Does the black hole shadow probe the event horizon geometry?,''
Phys.\ Rev.\ D \textbf{97}, no. 8, 084020 (2018).


\bibitem{bardeen} J. M. Bardeen. Gordon and Breach. in Black Holes
(Les Astres Occlus), C. Dewitt and B. S. Dewitt (eds.) pp. 215 - 239 (1973).

\bibitem{Chandra} S. Chandrasekhar. The Mathematical Theory of Black
Holes (Oxford University Press, New York) (1998).

%\cite{Hioki:2009na,Johannsen:2010ru,Nedkova:2013msa

\bibitem{Hioki:2009na}  K.~Hioki and K.~i.~Maeda,  
%``Measurement of the Kerr Spin Parameter by Observation of a Compact Object's Shadow,''
Phys.\ Rev.\ D \textbf{80}, 024042 (2009).

%\cite{Johannsen:2010ru}

\bibitem{Johannsen:2010ru}  T.~Johannsen and D.~Psaltis,  
%``Testing the No-Hair Theorem with Observations in the Electromagnetic Spectrum: II. Black-Hole Images,''
Astrophys.\ J.\ \textbf{718}, 446 (2010).

%\cite{Nedkova:2013msa}

\bibitem{Nedkova:2013msa}  P.~G.~Nedkova, V.~K.~Tinchev and S.~S.~Yazadjiev,
%``Shadow of a rotating traversable wormhole,''
Phys.\ Rev.\ D \textbf{88}, no. 12, 124019 (2013)

%\cite{Amarilla:2013sj}

\bibitem{Amarilla:2013sj}  L.~Amarilla and E.~F.~Eiroa,  
%``Shadow of a Kaluza-Klein rotating dilaton black hole,''
Phys.\ Rev.\ D \textbf{87}, no. 4, 044057 (2013).

%\cite{Abdujabbarov:2012bn}

\bibitem{Abdujabbarov:2012bn}  A.~Abdujabbarov, F.~Atamurotov, Y.~Kucukakca,
B.~Ahmedov and U.~Camci,  %``Shadow of Kerr-Taub-NUT black hole,''
Astrophys.\ Space Sci.\ \textbf{344}, 429 (2013).

\bibitem{Grenzebach:2014fha}  A.~Grenzebach, V.~Perlick and C. L\"{a}mmerzahl,  
%``Photon Regions and Shadows of Kerr-Newman-NUT Black Holes with a Cosmological Constant,''
Phys.\ Rev.\ D \textbf{89}, no. 12, 124004 (2014).

%\cite{Johannsen:2015hib}

\bibitem{Johannsen:2015hib}  T.~Johannsen \textit{et al.},  
%``Testing General Relativity with the Shadow Size of Sgr A*,''
Phys.\ Rev.\ Lett.\ \textbf{116}, no. 3, 031101 (2016).

%\cite{Giddings:2014ova}

\bibitem{Giddings:2014ova}  S.~B.~Giddings,  
%``Possible observational windows for quantum effects from black holes,''
Phys.\ Rev.\ D \textbf{90}, no. 12, 124033 (2014).

%\cite{Atamurotov:2013sca}

\bibitem{Atamurotov:2013sca}  F.~Atamurotov, A.~Abdujabbarov and B.~Ahmedov,
%``Shadow of rotating non-Kerr black hole,''
Phys.\ Rev.\ D \textbf{88}, no. 6, 064004 (2013). 


%\cite{Wei:2013kza}

\bibitem{Wei:2013kza}  S.~W.~Wei and Y.~X.~Liu,  
%``Observing the shadow of Einstein-Maxwell-Dilaton-Axion black hole,''
JCAP \textbf{1311}, 063 (2013).

%\cite{Sakai:2014pga}

\bibitem{Sakai:2014pga}  N.~Sakai, H.~Saida and T.~Tamaki,  
%``Gravastar Shadows,''
Phys.\ Rev.\ D \textbf{90}, no. 10, 104013 (2014).

%\cite{Perlick:2015vta}

\bibitem{Perlick:2015vta}  V.~Perlick, O.~Y.~Tsupko and
G.~S.~Bisnovatyi-Kogan,  
%``Influence of a plasma on the shadow of a spherically symmetric black hole,''
Phys.\ Rev.\ D \textbf{92}, no. 10, 104031 (2015).

%\cite{Abdujabbarov:2015xqa}

\bibitem{Abdujabbarov:2015xqa}  A.~A.~Abdujabbarov, L.~Rezzolla and
B.~J.~Ahmedov,  
%``A coordinate-independent characterization of a black hole shadow,''
Mon.\ Not.\ Roy.\ Astron.\ Soc.\ \textbf{454}, no. 3, 2423 (2015).

%\cite{Tinchev:2013nba}

\bibitem{Tinchev:2013nba}  V.~K.~Tinchev and S.~S.~Yazadjiev,  
%``Possible imprints of cosmic strings in the shadows of galactic black holes,''
Int.\ J.\ Mod.\ Phys.\ D \textbf{23}, 1450060 (2014).

%\cite{Wang:2018eui}

\bibitem{Wang:2018eui}  M.~Wang, S.~Chen and J.~Jing,  
%``Chaotic shadow of a non-Kerr rotating compact object with quadrupole mass moment,''
Phys. Rev. D \textbf{98}, no.10, 104040 (2018).

%\cite{Amarilla:2010zq}

\bibitem{Amarilla:2010zq}  L.~Amarilla, E.~F.~Eiroa and G.~Giribet,  
%``Null geodesics and shadow of a rotating black hole in extended Chern-Simons modified gravity,''
Phys.\ Rev.\ D \textbf{81}, 124045 (2010).

%\cite{Yumoto:2012kz}

\bibitem{Yumoto:2012kz}  A.~Yumoto, D.~Nitta, T.~Chiba and N.~Sugiyama,  
%``Shadows of Multi-Black Holes: Analytic Exploration,''
Phys.\ Rev.\ D \textbf{86}, 103001 (2012).



\bibitem{Takahashi:2005hy}  R.~Takahashi,  
%``Black hole shadows of charged spinning black holes,''
Publ.\ Astron.\ Soc.\ Jap.\ \textbf{57}, 273 (2005).



\bibitem{Papnoi:2014aaa}  U.~Papnoi, F.~Atamurotov, S.~G.~Ghosh and
B.~Ahmedov,  
%``Shadow of five-dimensional rotating Myers-Perry black hole,''
Phys.\ Rev.\ D \textbf{90}, no. 2, 024073 (2014).

\bibitem{Ovgun:2019jdo}
A.~\"{O}vg\"{u}n, I.~Sakalli, J.~Saavedra and C.~Leiva,
%``Shadow cast of non-commutative black holes in Rastall gravity,''
Mod. Phys. Lett. A \textbf{2050163}, 2020.


\bibitem{Dexter:2012fh}  J.~Dexter and P.~C.~Fragile,  
%``Tilted black hole accretion disc models of Sagittarius A*: time-variable millimetre to near-infrared emission,''
Mon.\ Not.\ Roy.\ Astron.\ Soc.\ \textbf{432}, 2252 (2013).


\bibitem{Moffat:2015kva}  J.~W.~Moffat,  
%``Modified Gravity Black Holes and their Observable Shadows,''
Eur.\ Phys.\ J.\ C \textbf{75}, no. 3, 130 (2015).



\bibitem{Younsi:2016azx}  Z.~Younsi, A.~Zhidenko, L.~Rezzolla, R.~Konoplya
and Y.~Mizuno,  
%``New method for shadow calculations: Application to parametrized axisymmetric black holes,''
Phys.\ Rev.\ D \textbf{94}, no. 8, 084025 (2016).



\bibitem{Johannsen:2015qca}  T.~Johannsen,  
%``Photon Rings around Kerr and Kerr-like Black Holes,''
Astrophys.\ J.\ \textbf{777}, 170 (2013).



\bibitem{Zakharov:2014lqa}  A.~F.~Zakharov,  
%``Constraints on a charge in the Reissner-Nordström metric for the black hole at the Galactic Center,''
Phys.\ Rev.\ D \textbf{90}, no. 6, 062007 (2014).



\bibitem{Cunha:2016bjh}  P.~V.~P.~Cunha, J.~Grover, C.~Herdeiro, E.~Radu,
H.~Runarsson and A.~Wittig,  
%``Chaotic lensing around boson stars and Kerr black holes with scalar hair,''
Phys.\ Rev.\ D \textbf{94}, no. 10, 104023 (2016).



\bibitem{Freivogel:2014lja}  B.~Freivogel, R.~Jefferson, L.~Kabir, B.~Mosk
and I.~S.~Yang,  %``Casting Shadows on Holographic Reconstruction,''
Phys.\ Rev.\ D \textbf{91}, no. 8, 086013 (2015).



\bibitem{Cunha:2016bpi}  P.~V.~P.~Cunha, C.~A.~R.~Herdeiro, E.~Radu and
H.~F.~Runarsson,  
%``Shadows of Kerr black holes with and without scalar hair,''
Int.\ J.\ Mod.\ Phys.\ D \textbf{25}, no. 09, 1641021 (2016).



\bibitem{Ohgami:2015nra}  T.~Ohgami and N.~Sakai,  %``Wormhole shadows,''
Phys.\ Rev.\ D \textbf{91}, no. 12, 124020 (2015).



\bibitem{Zakharov:2011zz}  A.~F.~Zakharov, F.~De Paolis, G.~Ingrosso and
A.~A.~Nucita,  
%``Shadows as a tool to evaluate black hole parameters and a dimension of spacetime,''
New Astron.\ Rev.\ \textbf{56}, 64 (2012).



\bibitem{Hennigar:2018hza}  R.~A.~Hennigar, M.~B.~J.~Poshteh and R.~B.~Mann,
%``Shadows, Signals, and Stability in Einsteinian Cubic Gravity,''
Phys.\ Rev.\ D \textbf{97}, no. 6, 064041 (2018).




\bibitem{Pu:2016qak}  H.~Y.~Pu, K.~Akiyama and K.~Asada,  
%``The Effects of Accretion Flow Dynamics on the Black Hole Shadow of Sagittarius A$^{*}$,''
Astrophys.\ J.\ \textbf{831}, no. 1, 4 (2016).



\bibitem{Sharif:2016znp}  M.~Sharif and S.~Iftikhar,  
%``Shadow of a Charged Rotating Non-Commutative Black Hole,''
Eur.\ Phys.\ J.\ C \textbf{76}, no. 11, 630 (2016).



\bibitem{Abdujabbarov:2016hnw}  A.~Abdujabbarov, M.~Amir, B.~Ahmedov and
S.~G.~Ghosh,  %``Shadow of rotating regular black holes,''
Phys.\ Rev.\ D \textbf{93}, no. 10, 104004 (2016).

\bibitem{Xu:2018mkl}
Z.~Xu, X.~Hou and J.~Wang,
%``Possibility of Identifying Matter around Rotating Black Hole with Black Hole Shadow,''
JCAP \textbf{10} (2018), 046

\bibitem{Gyulchev:2018fmd}  G.~Gyulchev, P.~Nedkova, V.~Tinchev and
S.~Yazadjiev,  %``On the shadow of rotating traversable wormholes,''
 Eur.\ Phys.\ J.\ C {\bf 78}, no. 7, 544 (2018).



\bibitem{Hou:2018bar}  X.~Hou, Z.~Xu, M.~Zhou and J.~Wang,  
%``Black Hole Shadow of Sgr A$^{*}$ in Dark Matter Halo,''
  JCAP {\bf 1807}, no. 07, 015 (2018).

\bibitem{Dokuchaev:2018kzk}
V.~Dokuchaev and N.~Nazarova,
%``Event horizon image within black hole shadow,''
J. Exp. Theor. Phys. \textbf{128} no.4, 578-585 (2019).

\bibitem{Mizuno:2018lxz}  Y.~Mizuno \textit{et al.},  
%``The Current Ability to Test Theories of Gravity with Black Hole Shadows,''
Nat.\ Astron.\  {\bf 2}, no. 7, 585 (2018).



\bibitem{Perlick:2018iye}  V.~Perlick, O.~Y.~Tsupko and
G.~S.~Bisnovatyi-Kogan,  
%``Black hole shadow in an expanding universe with a cosmological constant,''
Phys.\ Rev.\ D \textbf{97}, no. 10, 104062 (2018).



\bibitem{Stuchlik:2018qyz}  Z.~Stuchlak, D.~Charbulak and J.~Schee,  
%``Light escape cones in local reference frames of Kerr–de Sitter  black hole spacetimes and related black hole shadows,''
Eur.\ Phys.\ J.\ C \textbf{78}, no. 3, 180 (2018).



\bibitem{Shaikh:2018lcc}  R.~Shaikh, P.~Kocherlakota, R.~Narayan and
P.~S.~Joshi,  
%``Shadows of spherically symmetric black holes and naked singularities,''
Mon. Not. Roy. Astron. Soc. \textbf{482}, no.1, 52-64 (2019).



\bibitem{Eiroa:2017uuq}  E.~F.~Eiroa and C.~M.~Sendra,  
%``Shadow cast by rotating braneworld black holes with a cosmological constant,''
Eur.\ Phys.\ J.\ C \textbf{78}, no. 2, 91 (2018).


\bibitem{Mars:2017jkk}  M.~Mars, C.~F.~Paganini and M.~A.~Oancea,  
%``The fingerprints of black holes-shadows and their degeneracies,''
Class.\ Quant.\ Grav.\ \textbf{35}, no. 2, 025005 (2018).



\bibitem{Wang:2017hjl}  M.~Wang, S.~Chen and J.~Jing,  
%``Shadow casted by a Konoplya-Zhidenko rotating non-Kerr black hole,''
JCAP \textbf{1710}, no. 10, 051 (2017).



\bibitem{Tsukamoto:2017fxq}  N.~Tsukamoto,  
%``Black hole shadow in an asymptotically-flat, stationary, and axisymmetric spacetime: The Kerr-Newman and rotating regular black holes,''
Phys.\ Rev.\ D \textbf{97}, no. 6, 064021 (2018).

\bibitem{young} P. J. Young, Phys. Rev. D 14, 3281 (1976).



\bibitem{Singh:2017vfr}  B.~P.~Singh and S.~G.~Ghosh,  
%``Shadow of Schwarzschild–Tangherlini black holes,''
Annals Phys.\ \textbf{395}, 127 (2018).



\bibitem{Mureika:2016efo}  J.~R.~Mureika and G.~U.~Varieschi,  
%``Black hole shadows in fourth-order conformal Weyl gravity,''
Can.\ J.\ Phys.\ \textbf{95}, no. 12, 1299 (2017).



\bibitem{Huang:2016qnl}  Y.~Huang, S.~Chen and J.~Jing,  
%``Double shadow of a regular phantom black hole as photons couple to the Weyl tensor,''
Eur.\ Phys.\ J.\ C \textbf{76}, no. 11, 594 (2016).



\bibitem{Ghasemi-Nodehi:2016wao}  M.~Ghasemi-Nodehi and C.~Bambi,  
%``Note on a new parametrization for testing the Kerr metric,''
Eur.\ Phys.\ J.\ C \textbf{76}, no. 5, 290 (2016).



\bibitem{Tsukamoto:2014tja}  N.~Tsukamoto, Z.~Li and C.~Bambi,  
%``Constraining the spin and the deformation parameters from the black hole shadow,''
JCAP \textbf{1406}, 043 (2014).



\bibitem{Vincent:2016sjq}  F.~H.~Vincent, E.~Gourgoulhon, C.~Herdeiro and
E.~Radu,  %``Astrophysical imaging of Kerr black holes with scalar hair,''
Phys.\ Rev.\ D \textbf{94}, no. 8, 084045 (2016).

\bibitem{Younsi:2016azx}
Z.~Younsi, A.~Zhidenko, L.~Rezzolla, R.~Konoplya and Y.~Mizuno,
%``New method for shadow calculations: Application to parametrized axisymmetric black holes,''
Phys. Rev. D \textbf{94} no.8, 084025 (2016).

\bibitem{Konoplya:2020bxa}
R.~Konoplya and A.~Zinhailo,
%``Quasinormal modes, stability and shadows of a black hole in the novel 4D Einstein-Gauss-Bonnet gravity,''
arXiv:2003.01188 [gr-qc].

\bibitem{Konoplya:2019sns}
R.~Konoplya,
%``Shadow of a black hole surrounded by dark matter,''
Phys. Lett. B \textbf{795}, 1-6 (2019).

\bibitem{Konoplya:2019goy}
R.~Konoplya and A.~Zhidenko,
%``Analytical representation for metrics of scalarized Einstein-Maxwell black holes and their shadows,''
Phys. Rev. D \textbf{100} no.4, 044015 (2019).

\bibitem{Konoplya:2019fpy}
R.~A.~Konoplya, T.~Pappas and A.~Zhidenko,
%``Einstein-scalar–Gauss-Bonnet black holes: Analytical approximation for the metric and applications to calculations of shadows,''
Phys. Rev. D \textbf{101} no.4, 044054 (2020).

\bibitem{Contreras:2019cmf}
E.~Contreras, A.~Rincon, G.~Panotopoulos, P.~Bargueno and B.~Koch,
%``Black hole shadow of a rotating scale--dependent black hole,''
Phys. Rev. D \textbf{101} no.6, 064053 (2020).

\bibitem{Contreras:2019nih}
E.~Contreras, J.~Ramirez-Velasquez, A.~Rincon, G.~Panotopoulos and P.~Bargueno,
%``Black hole shadow of a rotating polytropic black hole by the Newman–Janis algorithm without complexification,''
Eur. Phys. J. C \textbf{79} no.9, 802 (2019).

\bibitem{Li:2020drn}
P.~Li, M.~Guo and B.~Chen,
%``Shadow of a Spinning Black Hole in an Expanding Universe,''
Phys. Rev. D \textbf{101} no.8, 084041 (2019).

\bibitem{Li:2019lsm}
C.~Li, S.~Yan, L.~Xue, X.~Ren, Y.~Cai, D.~A.~Easson, Y.~Yuan and H.~Zhao,
%``Testing the equivalence principle via the shadow of black holes,''
Phys. Rev. Res. \textbf{2}, 023164 (2020).

\bibitem{Bambi:2019tjh}
C.~Bambi, K.~Freese, S.~Vagnozzi and L.~Visinelli,
%``Testing the rotational nature of the supermassive object M87* from the circularity and size of its first image,''
Phys. Rev. D \textbf{100} no.4, 044057 (2019).

\bibitem{Psaltis:2020cte}
D.~Psaltis, L.~Medeiros, T.~R.~Lauer, C.~Chan and F.~Ozel,
%``Discretization and Filtering Effects on Black Hole Images Obtained with the Event Horizon Telescope,''
arXiv:2004.06210 [astro-ph.HE].

\bibitem{Johannsen:2012vz}
T.~Johannsen, D.~Psaltis, S.~Gillessen, D.~P.~Marrone, F.~Ozel, S.~S.~Doeleman and V.~L.~Fish,
%``Masses of Nearby Supermassive Black Holes with Very-Long Baseline Interferometry,''
Astrophys. J. \textbf{758}, 30 (2012).

\bibitem{Shaikh:2019fpu}
R.~Shaikh,
%``Black hole shadow in a general rotating spacetime obtained through Newman-Janis algorithm,''
Phys. Rev. D \textbf{100} no.2, 024028 (2019).

\bibitem{Vagnozzi:2020quf}
S.~Vagnozzi, C.~Bambi and L.~Visinelli,
%``Concerns regarding the use of black hole shadows as standard rulers,''
Class. Quant. Grav. \textbf{37} no.8, 087001 (2020).

\bibitem{Abdikamalov:2019ztb}
A.~B.~Abdikamalov, A.~A.~Abdujabbarov, D.~Ayzenberg, D.~Malafarina, C.~Bambi and B.~Ahmedov,
%``Black hole mimicker hiding in the shadow: Optical properties of the $\gamma$ metric,''
Phys. Rev. D \textbf{100} no.2, 024014 (2019).

\bibitem{Kumar:2019pjp}
R.~Kumar, S.~G.~Ghosh and A.~Wang,
%``Shadow cast and deflection of light by charged rotating regular black holes,''
Phys. Rev. D \textbf{100} no.12, 124024 (2019).

\bibitem{Amir:2017slq}
M.~Amir, B.~P.~Singh and S.~G.~Ghosh,
%``Shadows of rotating five-dimensional charged EMCS black holes,''
Eur. Phys. J. C \textbf{78} no.5, 399 (2018).

\bibitem{Bambi:2008jg}
C.~Bambi and K.~Freese,
%``Apparent shape of super-spinning black holes,''
Phys. Rev. D \textbf{79}, 043002 (2009).


\bibitem{Poplawski:2012bw}
N.~Poplawski,
%``Affine theory of gravitation,''
Gen. Rel. Grav. \textbf{46}, 1625 (2014).
  
  
  
  
  
  %%%
  \bibitem{Azreg-Ainou:2014pra}
M.~Azreg-Ainou,
%``Generating rotating regular black hole solutions without complexification,''
Phys. Rev. D \textbf{90} no.6, 064041 (2014).

\bibitem{myWald} R. M. Wald, \textit{General Relativity} (The University of Chicago Press, Chicago and London, 1984).

\bibitem{Lepe}
F.~Izaurieta and S.~Lepe,
%``Cosmological Dark Matter Amplification through Dark Torsion,''
arXiv:2004.06058 [gr-qc].


\end{thebibliography}
\end{document}